\documentclass[preprint,12pt]{aastex}

\newcommand{\degree}{\mbox{$^{\circ}$}}

\newcommand{\as}{\mbox{\arcsec}}

\newcommand{\vlsr}{$v_{LSR}$}

\def\lsim {$\rlap{\raise.4ex\hbox{$<$}}\lower.55ex\hbox{$\sim$}\,$}

\newcommand{\noprint}[1]{}

\newcommand{\msun}{\mbox{M$_\odot$}}

\newcommand{\mean}[1]{\mbox{$\langle#1\rangle$}} 
 
\newcommand{\ammonia}{\mbox{{\rm NH}$_3$}}
\newcommand{\coo}{$^{13}$CO}

\newcommand{\hcop}{HCO$^+$}
\newcommand{\hcopi}{H$^{13}$CO$^+$}

\newcommand{\nthp}{N$_2$H$^+$}
 
\input{epsf}

\newcommand{\rgal}{\mbox{$R_{gal}$}}

\begin{document}

               
\title {\bf The Bolocam Galactic Plane Survey. X. A Complete Spectroscopic Catalog of
Dense Molecular Gas Observed toward 1.1 mm Dust Continuum Sources with $7.5$\degree $\leq l \leq 194$\degree  }
\author{Yancy L. Shirley\altaffilmark{1,2},
	Timothy P. Ellsworth-Bowers\altaffilmark{3},
        Brian Svoboda\altaffilmark{1},
	 Wayne M. Schlingman\altaffilmark{3},
	Adam Ginsburg\altaffilmark{3},
	Erik Rosolowsky\altaffilmark{4},
       Thomas Gerner\altaffilmark{5}
	Steven Mairs\altaffilmark{6},
  	Cara Battersby\altaffilmark{3},
	Guy Stringfellow\altaffilmark{3},
	 Miranda K. Dunham\altaffilmark{7},
	Jason Glenn\altaffilmark{3},
	\and
        John Bally\altaffilmark{3}
	}
\altaffiltext{1}{Steward Observatory, 933 N Cherry Ave., Tucson, AZ 85721 USA}
\altaffiltext{2}{Adjunct Astronomer, The National Radio Astronomy Observatory}
\altaffiltext{3}{CASA, University of Colorado, CB 389, Boulder, CO 80309, USA}
\altaffiltext{4}{Department of Physics, University of
		 Alberta, 4-181 CCIS Edmonton AB T6G 2E1, Canada}
\altaffiltext{5}{Max-Planck-Institut f\"ur Astronomie (MPIA), Königstuhl 17, 69117, Heidelberg, Germany}
\altaffiltext{6}{Department of Physics and Astronomy, University of Victoria, P.O. Box 3055, STN CSC, Victoria, BC V8W 3P6, Canada}
\altaffiltext{7}{Department of Astronomy, Yale University, P.O. Box 208101, New Haven, CT 06520, USA}
 
\begin{abstract}
The Bolocam Galactic Plane Survey (BGPS) is a 1.1 mm continuum survey of dense
clumps of dust throughout the Galaxy covering 170 square degrees. 
We present spectroscopic observations using the Heinrich Hertz
Submillimeter Telescope of the dense gas tracers,
\hcop\ and \nthp\ $3-2$, for all 6194 sources in
the Bolocam Galactic Plane Survey v1.0.1 catalog between
$7.5$\degree $\leq l \leq 194$\degree .
This is the largest targeted spectroscopic survey of dense molecular gas
in the Milky Way to date.  We find unique velocities for 3126 ($50.5$\%) of the BGPS
v1.0.1 sources observed.  Strong \nthp\ $3-2$ emission ($T_{mb} > 0.5$ K) 
without \hcop\ $3-2$ emission does not occur in this catalog.
We characterize the properties of the 
dense molecular gas emission toward the entire sample.
\hcop\ is very sub-thermally populated and the
3-2 transitions are optically thick toward most BGPS clumps.
The median observed line width is $3.3$ km/s
consistent with supersonic turbulence within BGPS clumps. 
We find strong correlations between dense molecular gas 
integrated intensities and
$1.1$ mm peak flux and the gas kinetic temperature derived
from previously published \ammonia\ observations.  These intensity correlations
are driven by the sensitivity of the $3-2$ transitions to excitation conditions
rather than by variations in molecular column density or abundance.
We identify a subset of
113 sources with stronger \nthp\ than \hcop\ integrated intensity, but we find no
correlations between the \nthp /\hcop\ ratio and 1.1 mm continuum flux density, gas
kinetic temperature, or line width.   
Self-absorbed profiles are rare ($1.3$\%). 

\end{abstract}



\section{Introduction}

In the last few years, comprehensive surveys of the Milky Way galaxy
have been performed at submillimeter and millimeter wavelengths
that have imaged dust continuum emission and have provided 
the most complete census of embedded sites of star formation
in our Galaxy.  The Bolocam Galactic Plane Survey (BGPS) version 1.0 mapped over 170 square degrees of the Galactic plane at 1.1 mm (Aguirre et al. 2011)
and has catalogued over 8400 continuum sources (Rosolowsky et al. 2010).  
The BGPS observed the entire first quadrant in a strip that is 1\degree\
wide in galactic latitude with selected regions observed in the second
($l \approx$ 98\degree , 111\degree , 133\degree $-$ 136\degree ), 
third (188\degree $\leq l \leq$ 192\degree )
and fourth quadrants ($l > 350$\degree ).   A complementary survey at 870 $\mu$m,
ATLASGAL, has mapped sections of the southern Milky Way with a 2\degree\
wide strip in galactic latitude between longitudes of $300$\degree $\leq l \leq 60$\degree\
(Contreras et al. 2013).  Also, the \textit{Herschel Space Observatory} survey, Hi-Gal, 
has mapped the Galactic plane at 70, 160, 250, 350, and 500 $\mu$m (Molinari et al. 2012).  
When the complete ATLASGAL and Hi-Gal survey source catalogs are released and are combined
with the BGPS catalogs, 
the final source catalog will contain tens
of thousands of embedded sites of star formation in the Milky Way.

The population of objects discovered in the BGPS, ATLASGAL, and Hi-Gal surveys 
include dense starless and star-forming cores at the closest distances, 
clumps (unresolved collections of cores), and clouds at the farthest distances
(Dunham et al. 2011a).
In order to study the physical properties (e.g. size, mass, luminosity) 
of these embedded star-forming regions, their distances must first be determined.
Distances in the Galactic plane may be derived
using the Galactic rotation curve and a measurement of the
source $v_{LSR}$.  In the second and third quadrants, kinematic
distances are uniquely determined; but, in the first and fourth quadrants,
each measured $v_{LSR}$ corresponds to both a near and far kinematic distance
resulting in a kinematic distance ambiguity that must be resolved
using additional information.

Assignment of kinematic distances first and foremost requires a
determination of the unique $v_{LSR}$ of the 1.1 mm clump.   
CO is too easily excited in low density
molecular gas to be a unique kinematic probe ($n_{eff} < 10^2$ cm$^{-3}$)\footnote{The effective excitation density, $n_{eff}$, is defined as the density for which a $T_R = 1$ K line would
be observed (see Evans 1999 for assumptions and Table 1 of Reiter et al. 2011a).}.  
Along lines-of-sight that cross multiple arms of the Galaxy, as often happens
in the first quadrant of the Galaxy, CO $1-0$ emission displays 
complicated spectral shapes with multiple components 
spanning many tens of km/s.  
The gas directly
associated with the clumps is best probed
by a dense molecular gas tracer. 
Schlingman et al. (2011) performed a pilot survey of slightly less than 
1/3 of the BGPS clumps (1882 sources) in the first and second quadrant of the Galactic plane
in the dense gas tracers \hcop\ and \nthp\ $3-2$ ($n_{eff} \sim 10^4$ cm$^{-3}$).
They showed that \hcop\ and \nthp\ $3-2$ are unique kinematic tracers of BGPS clumps with less than
$2$\%\ of sources observed displaying multiple velocity components.
They also resolved the kinematic distance ambiguity of a small
subset ($N = 529$) of the BGPS clumps.  Schlingman et al. (2011) found a median radius  
of 0.75 pc, a median mass of $320$ \msun , a median volume density of 
2400 cm$^{-3}$, and a median gravitational free-fall time of 750,000 years for this
subset of sources. The resulting clump mass
distribution ($dN/d\log M$) had a slope of $-0.8$ that is intermediate between the
CO cloud mass distribution ($-0.6$; Solomon et al. 1987) and the Salpeter IMF ($-1.35$;
Salpeter 1955; Scalo 1986) and is consistent with simulations of the fragmentation
of turbulent compressible gas with a Kolmogorov spectrum (i.e. Hennebelle \& Falgarone 2012).
Schlingman et al. also observed a breakdown
in the size-linewidth in the dense molecular gas probed by \hcop , possibly due
to turbulent feedback on small scales from embedded protostellar sources
(see Murray et al. 2011).
While the Schlingman survey provided an initial look the physical properties
of BGPS clumps, the physical properties derived from the 
known distance sample in that paper represents less than 
$10$\%\ of the full BGPS v1.0.1 catalog and is biased toward the brightest
1.1 mm sources.

In this paper, we complete the spectroscopic observations started by
Schlingman et al. (2011) and
provide a complete spectroscopic catalog of observations of
dense molecular gas as traced by \hcop\ and \nthp\ $3-2$ for all 6194 sources
in the BGPS v1.0.1 catalog between $7.5$\degree $\leq l \leq 194$\degree .
We characterize the properties of the molecular emission 
($v_{LSR}$, I, $\Delta v$, etc.) 
for the spectroscopic catalog in this paper.  
A companion paper
develops a Bayesian method for resolving the kinematic distance ambiguity for
a subset of sources in the first quadrant by deriving 
Distance Probability Density Functions
(DPDFs, Ellsworth-Bowers et al. 2013). 
The observing and calibration
procedures are explained in \S2, detection statistics are described in \S3,
and the source $v_{LSR}$ are extracted in \S4.1.  The properties of the molecular
emission are analyzed in \S4.2-\S4.5.    

\section{Observations}

\subsection{HHT Setup and Calibration}

4705 BGPS clumps in the range $7.5$\degree $\leq l \leq 194$\degree\
were observed during $51$ observing shifts between February 28, 2011
and December 8, 2012 with the 10m Heinrich Hertz Telescope (HHT)\footnote{The
Heinrich Hertz Submillimeter Telescope is operated by the Arizona Radio Observatory}.
The source catalog was determined by observing all Bolocat 
v1.0.1 sources that were not previously observed by Schlingman et al. (2011)
plus re-observing sources where the pointing in Schlingman et al.
exceeded half the HHT beamwidth ($15$\as ) from the v1.0.1 catalog
peak 1.1 mm continuum position.  Re-observation of a subset (393) of the
Schlingman sources was necessary because the Schlingman et al.
observations did not have access to the final v1.0.1 catalog
positions, but were instead observed with an earlier (v0.7)
version of the Bolocat (see \S2.2).  The observations
in this paper are ultimately combined with the Schlingman et al. observations
for sources with good correspondence to the v1.0.1 positions
($\Delta \theta < 15$\as ).  The resulting spectroscopic
catalog is a complete set of observations of every source (6194)
in the v1.0.1 Bolocat with $7.5$\degree $\leq l \leq 194$\degree .

The HHT observational procedure was similar to 
that used to observe the initial sample of 1882
BGPS clumps in Schlingman et al. (2011).   Observations were
performed with the 1mm prototype-ALMA dual polarization, sideband-separating receiver.  
The receiver was tuned to \hcop\ $3-2$ (267.5576259 GHz)
in the lower sideband (LSB) and to \nthp\ $3-2$  (279.5118379 GHz) in the upper sideband (USB).
Rejection between the sidebands was measured by observing the
bright Galactic source, W75(OH), for $10$ minutes each observing shift.  
The median rejections for the upper sideband were $-15.7$ dB and $-8.6$ dB for vertical
and horizontal polarization respectively.  The 4-IF output was
connected to one set of filterbanks with 1 MHz spectral resolution and
512 MHz bandwidth.  This spectral resolution corresponds to a velocity resolution 
of $1.12$ km/s in the LSB and $1.07$ km/s in the USB.
Each source was observed in
position-switching mode with 1 minute of ON-source integration time.
Emission free OFF positions for every half-degree of the galactic plane
were determined by Schlingman et al. (2011).

Observations at the HHT are placed on the $T_A^*$ scale
using the standard Chopper-Wheel Calibration method (Penzias \& Burrus 1973).
Observations of planets are then used to determine the main beam
efficiency and put the spectra on a final $T_{mb}^{pol} = T_A^*/\eta_{mb}\eta_{pol}$
scale, where $\eta_{mb}\eta_{pol}$ is the product of the telescope main beam
efficiency and the coupling efficiency between the receiver optics for each
polarization and the telescope.  
Unfortunately, during the bulk of the observing  
(47 out of 51 shifts), the only planet available for observation 
was Saturn.  Saturn has two potential problems for calibration.  First, 
the rings contribute to the 1.1 mm continuum and also block continuum
from the planet.  Fortunately though, the rings
were oriented with inclinations between $7.3$\degree\ and $9.8$\degree\
which correspond to the shallow minimum in the
millimeter brightness temperature (Weiland et al. 2011).  Second, Saturn also has pressure broadened 
PH$_3$ $J_K = 1_0 - 0_0$ emission at 267 GHz in the atmosphere (Encrenaz \& Moreno 2002).
The lower sideband (\hcop\ $3-2$) calibration is strongly affected
by this absorption and cannot be used to calibrate the main beam
efficiency.  The upper sideband (\nthp\ $3-2$) is much less affected
by PH$_3$ absorption and was used to calibrate the upper sideband
main beam efficiencies.  Jupiter was available for observations only
after June, 2011 and the ratio between lower sideband and upper sideband
efficiencies was applied to the Saturn calibration to derive
the lower sideband main beam efficiencies.

As a consistency check, we also observed the source W75(OH) during every
shift.  The integrated intensity of W75(OH) correlates with the main
beam efficiency measured using Saturn (Figure 1).
A significant jump in the calibration is seen for both
Saturn and W75(OH) at MJD = 55675 days.  This discontinuity is due to a
warmup of the receiver and subsequent adjustment of the feedhorn optics.
The increase in efficiencies at this date occurs for all polarizations and sidebands
simultaneously.  The main beam efficiencies used in this paper are listed in Table 1.  
The final calibration numbers agree well with those used in Schlingman et al.
The final spectrum is the baseline rms weighted-average of 
the calibrated vertical and horizontal polarizations
$T_{mb} = (T_{mb}^{Vpol}/\sigma_{T^{Vpol}_{mb}}^2 + T_{mb}^{Hpol}/\sigma_{T^{Hpol}_{mb}}^2) / 
( 1/\sigma_{T^{Vpol}_{mb}}^2 + 1/\sigma_{T^{Hpol}_{mb}}^2))$ .

\subsection{A Note About BGPS Catalog Versions}

A spectroscopic survey of a large number of sources, such as the one presented 
in this paper, requires using a fixed source catalog during the course of
the survey.  At the start of the Schlingman et al. (2011) observations in November 2008,
the final BGPS v1.0 source catalog had not been completed, so Schlingman et al.
utilized the preliminary Bolocat v0.7 catalog.  The results from 
those early spectroscopic observations helped tune the inputs for the seeded-watershed algorithm
ultimately used to generate the v1.0 catalog (Rosolowsky et al. 2010).  At the beginning
of this current survey in February 2011, the version 1.0.1 catalog was the
latest version of the BGPS source catalog available and forms the basis for
this paper.  In February 2013, a re-reduction of the BGPS (version 2.0) became
available (Ginsburg et al. 2013).  
Simulations and testing of the new pipeline reduction
plus characterization of the BGPS v2.0 angular transfer function are 
reported in Ginsburg et al. (2013).
The resulting v2.0 BGPS maps recover
larger scale structure better than v1.0 maps.  A new v2.0 source catalog using the same inputs
for the seeded-watershed algorithm that was used to generate the 
v1.0.1 catalog is also now available; however, the source selection algorithm
has not been optimized for the noise structure in the new v2.0 maps. 
The fluxes quoted throughout this paper are derived from the new v2.0 BGPS 1.1 mm images. 
Since these new data products only became available after 
spectroscopic observations were completed (December 2012), we shall reference all spectroscopic
observations sequentially in this paper to the v1.0.1 source names and catalog numbers  
in the spectral properties table (Tables 2 and 3).

\section{Detection Statistics}

The complete spectroscopic survey reaches a mean baseline
rms of $58$ mK for spectra in the LSB and $62$ mK for
spectra in the USB.  As a result, the mean $3\sigma_{T_{mb}}$ limits 
for \hcop\ $3-2$ and \nthp\ $3-2$ in this survey are $0.174$ K and $0.186$ K
respectively.  During the best weather, baseline rms of $30$ mK were
observed.  The distribution of $\sigma_{T_{mb}}$ with Galactic longitude 
is shown in Figure 2.

Detections are confirmed
visually, independently in each polarization, before the final spectrum is flagged.
We adopt the flagging scheme in Schlingman et al. (2011)
with two changes.  We have eliminated the flag for large
line wings (flag = $4$ in the Schlingman et al. scheme);
sources in the Schlingman catalog with a flag of 4 have been
re-classified with a flag of $1$ (single peak detection) or $3$ (self-absorbed profile) 
in the final catalog presented in this paper.   Also, to eliminate
numerical gaps in our flag scheme, we have re-classified all self-absorbed
profiles in the Schlingman et al. catalog (originally a flag of $5$) as a flag of 
$3$.  The flags in the merged catalog span from $0$ to $3$ and correspond to:
a source with a flag = 1 indicates a single $v_{LSR}$
component detected at the $> 3\sigma_{T_{mb}}$ level; a source with a flag = 2
indicates a multiple $v_{LSR}$ components detected; a source with a flag = 3 
indicates a source with possible self-absorption; and a source with a flag = 0
indicates a non-detection.

Examples of spectra with different flag combinations are shown in Figure 3.  
The vast majority of detections are similar to source 3054 (G23.370+0.140) 
with a single Gaussian
peak in both \hcop\ and \nthp\ $3-2$ (\hcop\ flag = $1$ and \nthp\ flag = $1$).
Source 1315 (G07.636-0.150) is an example of a \nthp\ detection that is stronger than
a \hcop\ detection while source 6620 (G78.233-0.347) is an example of a \hcop\ detection
and a \nthp\ non-detection.  Source 2060 (G14.540-0.210) is an example of a non-detection
in both lines over the full velocity range permissible for objects at $l = 14$\degree\ 
($-60$ to $+160$ km/s).
The four panels on the right of Figure 3 display examples of detections with multiple
velocity peaks.  Source 3244 (G24.143+0.128) shows two clearly separated 
velocity peaks ($\Delta v_{LSR} = 60$ km/s) indicative of two physically 
separate dense clumps along the
same line-of-sight.  Occasionally, the two peaks are blended.  In the case of source
2162 (G15.079-0.604), the \nthp\ $3-2$ spectrum peaks at the same velocity as one of the
\hcop\ $3-2$ peaks; therefore, we do not classify this source as a self-absorbed profile, but
as multiple velocity components in \hcop\ $3-2$ (\hcop\ flag = $2$).  
The two panels on the bottom right of
Figure 3 show examples of blue-skewed and red-skewed 
self-absorbed \hcop\ $3-2$ profiles where the \nthp\ $3-2$ emission
does peak near the self-absorption minima (\hcop\ flag = $3$).

The detection fractions for each spectroscopic flag are summarized in pie charts in
Figure 4.  \hcop\ $3-2$ emission was detected toward a 
total of $3206$ ($51.8$\%) sources.  \nthp\ $3-2$ emission was detected toward a smaller
fraction, $1878$ ($30.3$\%) sources.  Multiple velocity
components are rare in these dense gas tracers
with $84$ \hcop\ $3-2$ multiple velocity detections ($1.3$\%)
and $32$ \nthp\ $3-2$ multiple velocity detections ($0.5$\%).
There are only $4$ \nthp\ $3-2$ detections ($0.06$\%) that lack a corresponding
\hcop\ $3-2$ detection.  None of
these four unique \nthp\ $3-2$ detections are very significant, only ranging from 
$T_{mb}^{pk} = (3 - 5)\sigma_{T_{mb}}$.
As an example, source 1705 (G012.387+00.230) has a $3.7\sigma$ detection
in $T_{mb}$ for \nthp\ $3-2$ but there is a potential \hcop\ 3-2 detection at 
only $2.9\sigma$ which falls just short of the criteria set in this paper 
for a \hcop\ detection flag = 1.
Strong \nthp\ $3-2$ emission without an \hcop\ $3-2$ detection does not occur for
the BGPS sources observed in this spectroscopic catalog.
Therefore, we have observed a unique velocity component
for $3126$ sources ($50.5$\%) in the Bolocat v1.0.1
within the range $7.5$\degree $\leq l \leq 194$\degree .

The distributions of $T_{mb}$ for both tracers are shown in Figure 5.
Both histograms peak just above the $3 \sigma_{T_{mb}}$ limit indicating
that deeper integration is likely to detect more sources in
the lines of \hcop\ and \nthp\ $3-2$.  The total observing time for the complete
spectroscopic survey is $> 600$ hours; a modest decrease of a
factor of two in the rms noise level would require four times longer than
the current survey.
The detection statistics are a function of Galactic longitude 
in this survey (see Figure 2). 
The detection fraction does not strongly correlate with the baseline 
rms in the survey (Figure 5).
For instance, the region near $l = 30$\degree\ which has the
largest number of 1.1 mm sources within a $1$\degree\ bin has a detection
fraction of only $30$\%\ despite having small baseline rms compared
to the  entire survey ($\sigma_{T_{mb}} = 40$ mK).  An opposite example
can be found at $l = 83$\degree\ where a higher baseline rms results in a 
lower detection fraction; however, these instances are rare in this survey.

In order to understand this result, we must
first compare the detection statistics to the flux density of BGPS
sources.  The detection fraction is a strong
function of the $1.1$mm flux density of BGPS sources (Figure 4).  v1.0.1 sources
with $1.1$mm flux densities below 200 mJy have a less than $50$\%\
detection percentage with the detection fraction decreasing rapidly
below this flux density.  This flux density corresponds to 
a column density limit of $N_{H_2} = 4 \times 10^{21}$ cm$^{-2}$
assuming a dust temperature of 20 K, gas mass to dust mass ratio of 100:1,
and Ossenkopf \& Henning opacities ($\kappa_{1.1}^{\rm{OH5}} = 1.14$ cm$^2$ g$^{-1}$;
Ossenkopf \& Henning 1994).
In contrast, sources with flux densities
above 500 mJy have a detection fraction approaching $90$\%\ 
(N$_{H_2} \geq 1 \times 10^{22}$ cm$^{-2}$).  
These variations in detection fraction with Galactic longitude primarily
represent variations in the number of low flux density sources
selected by the Bolocat seeded-watershed algorithms in different,
sometimes crowded, regions of the Galactic plane (see Figure 17 of Rosolowsky et al. 2010).  
The region toward $l = 30$\degree\ contains a larger fraction of low
flux density ($S_{1.1} < 200$ mJy) sources than other regions in 
the Galactic plane resulting in the lower overall detection fraction.
This $l=30$\degree\ region of the Galactic plane is
crowded and confused as the observed line-of-sight crosses the
Sagittarius arm, the tangent of the end of the molecular bar, Perseus arm,
and an outer Galactic arm.  

We can test the fidelity of low flux density sources in the v1.0.1 
catalog by directly comparing to the detection statistics
with the new v2.0 catalog.  For sources with no v2.0 catalog
counterpart within $60$\as\ of the v1.0.1 source position (1813 sources), 
the \hcop\ $3-2$ non-detection fraction (\hcop\ flag = 0) increases from $48.0$\%\ for
the complete v1.0.1 catalog to $73.7$\%\ for this subset. 
Nearly half of all v1.0.1 \hcop\ $3-2$ non-detections are no longer directly 
associated with sources in the v2.0 catalog.  Furthermore,
those v1.0.1 sources that lack a v2.0 counterpart are predominantly 
low flux density v1.0.1 sources (Ginsburg et al., 2013)
A significant fraction of low flux density v1.0.1 sources may not
be real or are low volume density sources at the limit of statistical significance
in the 1.1 mm map.
Conversely, there are 476 v1.0.1 sources which have \hcop\ $3-2$ detections
(\hcop\ flag = 1 or 3)
but no nearby ($< 60$\as ) counterpart in the v2.0 catalog.
These results indicate that the v2.0 catalog has a better chance
of selecting clumps with detectable dense molecular gas (at the limit
of this spectroscopic survey) but that the v2.0 catalog is incomplete and misses 
a significant number of sources
with dense molecular gas in the BGPS survey.  This comparison only serves to highlight
the difficulties inherent in source selection algorithms and also
how surveys of dense molecular gas may be used to test the fidelity of
those source catalogs.

\section{Properties of Molecular Detections}

In this section, we analyze the properties of the
\hcop\ and \nthp\ $3-2$ emission for the complete set of $3210$ 
spectroscopic detections.  The spectra are analyzed using the same analysis
techniques presented in Schlingman et al. (2011).  Namely,
a single component Gaussian is fit to \hcop\ $3-2$ spectra while a multiple-component
hyperfine fit is performed on \nthp\ $3-2$ spectra.  In cases where a single component
Gaussian fit fails or is not appropriate to described the observed line shape,
the velocity of $T_{mb}^{pk}$ is reported.
The integrated
intensities ($I = \int T_{mb} dv$) are determined from direct integration of the spectra.
Tables 2 and 3 list the derived spectroscopic properties ($\sigma_{T_{mb}}$, $v_{LSR}$, $T_{mb}^{pk}$,
I, FWHM, and FWZI) for \hcop\ and \nthp\ $3-2$ observations of all 6194 sources in the 
v1.0.1 catalog with $7.5$\degree $\leq l \leq 194$\degree .

In addition to the molecular lines observed in this survey, 
NH$_3$ is another popular dense gas tracer because it is
easily excited in dense  
molecular gas and the ratio of the (1,1) and (2,2) inversion lines may be
used to determine the gas kinetic temperature (Ho \& Townes 1983).  
There have been two substantial Galactic NH$_3$ surveys published in 
the past two years.  The Dunham et al. (2011) survey
used the 100m Green Bank Telescope to follow-up a 
sample of 631 BGPS sources in the (1,1), (2,2), and (3,3) inversion
lines.  The Wienen et al. (2012)
survey used the 100m Effelsberg Telescope (of which the
inner 80m is practical for 1 cm NH$_3$ observations) to
follow-up 862 ATLASGAL sources in the (1,1) and (2,2) inversion lines.  
We have directly compared the positions observed in the Dunham et al. and
Wienen et al. NH$_3$ catalogs with the HHT pointing
positions to find a subset of 546 sources with NH$_3$
pointings within $15$\as\ and having both published 
NH$_3$ detections and \hcop\ $3-2$ detections.  The distribution of
gas kinetic temperatures for these 546 sources is shown in
Figure 5.  The median is $T_k = 18.3$ K with a positively skewed tail of
kinetic temperatures.
We shall also use the gas kinetic temperature from 
this overlap subset of 546 sources in the subsequent 
analysis.

\subsection{Kinematics of BGPS Sources}

The primary purpose of this spectroscopic survey is to find a
unique $v_{LSR}$ for each BGPS source.  The $v_{LSR}$ for
molecular detections is determined from a Gaussian fit to
the \hcop\ $3-2$ line for sources with a flag = $1$.  Observations of
\nthp\ $3-2$, when detected, provide an independent measurement of
the velocity of dense molecular gas.

Figure 6 plots the distribution of 3126 observed velocities
with Galactic longitude overlaid on the CO $1-0$ integrated
intensities map from the Dame et al. (2001) survey.
The BGPS clumps generally follow the regions of strong CO $1-0$ emission and
trace major kinematic features in the Galaxy.  The most
prominent concentration of sources exists at $l \sim 30$\degree\
and $v_{LSR} \sim +90$ km/s corresponding to the tangent to
the molecular ring and the central molecular bar.  This also corresponds to
a secondary maximum in the Galactic longitude distribution of
sources in the first quadrant (Aguirre et al. 2011, Beuther et al. 2012)
with the peak of that distribution being toward $l = 0$\degree .
The second most prominent kinematic feature is the molecular
ring which extends diagonally in Figure 6 from $l = 30$\degree\ to
the longitude limit in this survey ($l = 7.5$\degree ).
The vast majority of sources between $l = 7.5$\degree\ and $l = 30$\degree\
are associated with this structure in the Milky Way.  Unfortunately,
these sources suffer from the kinematic distance ambiguity within
the molecular ring which becomes more severe as Galactic longitude
decreases from the tangent point with the bar near $l = 30$\degree .
BGPS clumps can be observed on the far side of the Galaxy as
evinced by the sources tracing the CO-delineated outer extension of the Norma arm
from $l = 25$\degree\ to $45$\degree\ and $v_{LSR} = -10$ to
$-50$ km/s.  Since the BGPS is confined to $\pm 0.5$\degree\ for
most longitude ranges, the BGPS has difficulty tracing sources in the newly discovered
outermost arm of the Milky Way (Dame \& Thaddeus 2011) because the
Galactic warp projects sources to Galactic latitudes greater than the 
typical $0.5$\degree\ extent of the BGPS.
A secondary clump of sources is distinct toward $l \sim 80$\degree\ and
$v_{LSR} \sim 0$ km/s that are associated with the famous Cygnus OB associations
and star-forming regions within the local arm (Reipurth \& Schneider 2008).

The $v_{LSR}$ of sources with both \hcop\ and \nthp\ $3-2$ detections
are shown in Figure 6.  The velocities of these two tracers are very well correlated
and centered on the one-to-one line.  Since the \nthp\ $3-2$ spectra were reduced
independently of the \hcop\ detections, this result indicates high fidelity
of the $v_{LSR}$ determined from the dense gas detections reported in this paper.

We also directly compare the $v_{LSR}$ from our survey with the spectra
that were observed within $15$\degree\ $< l < 56$\degree\ by the
\coo\ $1-0$ Galactic Ring Survey (GRS; Jackson et al. 2006, Roman-Duval et al. 2009).
GRS spectra are obtained for the v1.0.1 source positions from which
the velocity of the peak $T_{mb}$ were calculated.  1681 sources have both
\coo\ $1-0$ GRS spectra and a BGPS dense gas detection.
The peak \coo\ $1-0$ velocities agree within $2$ km/s with BGPS velocities 
for $81.4$\% of the sources.  There is a substantial number of sources with discordant velocities:  
217 sources ($12.9$\%) have a velocity offset $> 7$ km/s.
As can be seen in the bottom right panel of Figure 6, many of these
large discrepancies are tens of km/s and it would be unwise to blindly 
use the \coo\ $1-0$ peak velocity to determine kinematic distances.

The comparison statistics improve slightly for the small subset of $141$ sources
observed by Eden et al. (2012) in the \coo\ $3-2$ transition
with $87.1$\% of sources with velocities that agree within $2$ km/s.  While the higher
excitation CO transition does a slightly better job of picking the unique 
velocity of dense gas, $^{12}$CO and \coo\ cannot be used to uniquely determine
the velocities of all BGPS clumps.  Observations in dense gas tracers (with a contamination
rate from multiple components of only $1.3$\%) are required
for certainty in determining $v_{LSR}$ of a BGPS clump.

Many BPGS sources appear to be physically associated with complexes of sources.  
This is evident in the kinematic finder chart centered on  $l = 10.75$\degree , 
$b = -0.3$\degree\
(Figure 7).  
The BGPS v2.0 1.1 mm continuum images is displayed in greyscale with flux density
indicated along the top.
Sources with unique velocity detections are displayed as green circles with the $v_{LSR}$
indicated above the source.  Red crosses correspond to positions with no detection.
Yellow diamonds correspond to sources with multiple velocity components.
The brightest 1.1 mm complex in the southwest quadrant of
Figure 7 ranges in velocity from $-4.1$ to
$-1.5$ km/s and appears to be both spatially and kinematically connected.
There is a second, weaker 1.1 mm complex to the northeast that ranges in velocity from
$+28.6$ to $+30.8$ km/s that may be two vertical (aligned in Galactic latitude) filaments.
Two sources at $-5.3$ and $+51.0$ km/s are clearly interlopers and are not associated once
$v_{LSR}$ is taken into account.  This example highlights the need for kinematic
information to determine whether objects might be associated.

A simple tool for analyzing whether sources are kinematically associated is to study 
the distribution of nearest neighbors to each spectroscopically-detected source.
We calculate the difference in velocity, $\Delta v_{LSR}$, for the nearest
neighbor for all single component molecular detections.  The distribution
of $\Delta v_{LSR}$ is shown in Figure 8 for all nearest neighbor sources
that are within $10$\arcmin\ (1516 sources).  Just over 2/3 of nearest neighbors
appear kinematically associated ($69$\% of nearest neighbors have
$\Delta v_{LSR} < 7$ km/s).  This is not a surprising result because multiple
BGPS clumps may lie within the same molecular cloud traced by CO (e.g. Eden et al. 2012, 2013).
Identifying physically connected clumps or filaments is difficult in the BGPS maps because of
the spatial filtering due to atmospheric subtraction; however, observations from the 
\textit{Herschel Space Observatory} will suffer less severe spatial filtering
than ground-based observations.  Early \textit{Herschel} observations
revealed a plethora of filaments within the Galactic plane (e.g., Men'shchikov 2010,
Molinari et al. 2010, Arzoumanian et al. 2011, Palmeirim et al. 2012)
The kinematic information presented in this paper will be crucial for finding 
spatially and kinematically connected filaments in
observations of the Galactic plane from the \textit{Herschel Space Observatory}.

\subsubsection{A Methodology for Deriving Heliocentric Distances}

Ultimately, in order to derive physical properties of BGPS clumps
such as size, mass, and luminosity, we must determine the heliocentric distances, 
$d_{\odot}$.
For sources in the first quadrant, resolving the kinematic distance ambiguity (KDA)
presents a formidable problem.  The top left panel of Figure 8 shows the distribution
of the difference between the far and near heliocentric distance.  The majority of
sources observed in the first quadrant have $\Delta d_{\odot} = d_{\rm{far}} - d_{\rm{near}}$ that vary between
$3 - 10$ kpc.  For physical quantities, such as the mass or the luminosity,
that functionally depend on $d_{\odot}^2$, the resulting uncertainty are
factors of one to two orders of magnitude.

It is unlikely that we will be able to uniquely 
resolve the KDA for every spectroscopic detection in this catalog.
A statistical approach is more appropriate to characterize the probability of
finding a source at a given distance.  The companion paper to this
survey by Ellsworth-Bowers et al. (2013) develops a Bayesian technique for
deriving the posterior Distance Probability Density Functions (DPDF) for
BGPS clumps.  In this framework, the posterior distribution is given
by the product of the likelihood function (derived from $v_{LSR}$
and the rotation curve of the Galaxy) and prior distributions
that constrain the probability of finding the source at near or
far kinematic distances
\begin{equation}
DPDF(d_{\odot}) = \mathcal{L}(v_{LSR},l,b;d_{\odot})\, \prod_i P_i(d_{\odot},l,b) \;\; .
\end{equation}
The likelihood function, $\mathcal{L}(v_{LSR},l,b;d_{\odot})$, 
is a bimodal distribution with equal probability centered
on the near and far kinematic distances derived from the Reid et al. (2009) rotation 
model of the Galaxy with a $7$ km/s uncertainty (see Ellsworth-Bowers et al. 2012, Reid et al. 2009).
Ellsworth-Bowers et al. (2013) develops two prior distributions based on 
the azimuthally averaged distribution of H$_2$ in the Galaxy (Wolfire et al. 2003) 
and a radiative transfer model of the Galactic 8 $\mu$m emission 
(Robitaille et al. 2012) to derive the DPDF for BGPS sources
associated with 8 $\mu$m absorption features (EMAFs) observed in \textit{Spitzer} images.
This study shows that not all EMAFs can be automatically associated with
the near kinematic distance; 15\% of BGPS sources associated with EMAFs are placed
at or beyond the tangent distance. Additional prior
distributions (i.e. based on the presence or lack of HI absorption features;
see Roman-Duval et al. 2009) may be used to further constrain the DPDFs.
The distribution of a source property (i.e. size, mass, luminosity, etc.) 
may then be calculated using Monte Carlo
techniques that marginalize over distance (randomly drawn from the  DPDFs) 
and other relevant physical variables (i.e. dust temperature and dust opacity 
in the calculation of mass from the observed 1.1 mm flux; see Schlingman et al 2011).  
Calculation of DPDFs for the entire spectroscopic catalog is beyond the scope
of this current paper and is the subject of ongoing work by the BGPS team.

There is no distance ambiguity for Galactocentric radius, \rgal .  This
permits us to compare the molecular derived properties vs. \rgal\ for
the subset of 3126 unique kinematic detections.  The Galactocentric
radius may be derived from the observed $v_{LSR}$, $l$, $b$,
and the assumed rotation curve of the Galaxy ($v(r)$).  
We use the Reid et al. (2009) model for the rotation curve of the Milky Way
determined from parallax measurements of maser sources.  A \textit{FORTRAN}
program supplied by M. Reid (2009; private communication) 
calculates the near or far kinematic 
distances from observed $v_{LSR}$ and galactic coordinates.  
The bottom panel of Figure 8 shows the distribution 
of sources with \rgal .  Four distinct peaks are visible: 
the molecular ring at $4.5$ kpc, the Sagittarius arm
at $6.5$ kpc, the Local arm at $8.5$ kpc, and the outer Perseus arm
at $10.5$ kpc.  

\subsection{Molecular Intensity Comparisons}

The brightest \hcop\ $3-2$ sources have integrated
intensities over 150 K km/s.  The brightest \hcop\ $3-2$
source, (6362 G49.489-0.370) is also the brightest 1.1 mm source
and appears associated with the luminous infrared source 
W51 IRS2S (Wynn-Williams, Becklin, \& Neugebauer 1974).
Both the \hcop\ and \nthp\ $3-2$ spectra display a strong
red asymmetry likely indicative of the strong molecular
outflows in this region.  The brightest \nthp\ $3-2$ source
is 2152 (G015.013-00.674) with $I = 42.6$ K km/s and
is associated with the M17-SW region.

Schlingman et al. (2011) reported a correlation between the 
\hcop\ and \nthp\ $3-2$ integrated intensities and the
1.1m flux density for the subset of 1882 sources observed.
The correlations for the complete spectroscopic catalog
are plotted in Figure 9.  Similar positive correlations
are observed for both molecules (Pearson's product-moment correlation coefficient is
$r_{corr} = 0.78$ for \hcop\ $3-2$ and $r_{corr} = 0.75$ for \nthp\ $3-2$).  
The excellent overlap in points
between new observations presented in this paper the Schlingman catalog
indicates that the calibration between
the two sets of observations is consistent (\S2).

The bottom panels of Figure 9 also show 
a correlation
between \hcop\ integrated intensity and
kinetic temperature indicating that warmer sources
have more \hcop\ $3-2$ emission ($r_{corr} = 0.62$).  
\nthp\ $3-2$ emission is slightly less well
correlated with $T_k$ ($r_{corr} = 0.53$).  
There is, however, a tendency for \nthp\ $3-2$ emission to be brighter
in warmer clumps.  This is likely due to excitation
conditions in those clumps.  The $J = 3$ level is 
$26.8$ K above ground and the level populations
and subsequent intensity of the $3-2$ line are sensitive
to the gas kinetic temperature (see \S4.4).
It will be important to ultimately compare the $1-0$ transitions,
for instance from the MALT90 survey (Foster et al. 2011, 2013), to derive
the chemical abundances of \hcop\ and \nthp\
with less sensitivity to excitation conditions.

We also find positive correlations in the integrated intensities of 
\hcop\ and \nthp\ $3-2$ emission ($r_{corr} = 0.82$) 
and in the ratio $I_{\rm{line}}/S_{1.1}$ 
($r_{corr} = 0.62$; Figure 9).   This ratio of integrated intensity to 
1.1 mm flux density is linearly proportional to the molecular abundance 
if molecular emission is in the optically thin limit.  
We must caution that the assumption
that \hcop\ and \nthp\ $3-2$ emission is optically thin is likely untrue for most
BGPS sources and that optical
depth effects will modify optically-thin column densities by the factor
$\frac{\tau}{1 - exp(-\tau)} \sim \tau$ for thick lines (see \S4.4).  

\hcop\ and \nthp\ have opposite chemical behavior with respect to CO.
\hcop\ is primarily formed in the gas phase from reactions of CO with
H$_3^+$ while \nthp\ is destroyed by gas phase CO (see J{\o}rgensen et al. 2004).  
Since CO is adsorbed onto dust grains at high densities ($n > 10^4$ cm$^3$) and low 
temperatures ($T_k < 20$ K), the gas phase abundances of \hcop\
and \nthp\ are expected to be anti-correlated in cold, dense, heavily
CO-depleted environments.  
The correlations observed above seem to contradict the simple chemical
expectation if they are naively interpreted as representing variations
in the abundance of \hcop\ and \nthp .  However, we must be careful when
directly comparing integrated intensities since this quantity is function of
both the column density (or abundance) and the excitation conditions,
which play an important role for the $3-2$ transitions (see \S4.4).

A large ratio of \nthp /\hcop\ emission
may be a better indicator of a significant reservoir of cold, dense gas within
a BGPS clump.  
There are 113 sources (1.8\%) with brighter \nthp\ $3-2$
integrated intensities than \hcop\ $3-2$ lines.  
In contrast, the CHaMP survey (Census of High- and Medium-mass Protostars; 
Barnes et al. 2010, 2011) does not find any high-mass clumps in the their follow-up
mapping survey of 303 sources in the fourth quadrant with \nthp\ $1-0$ integrated
intensities larger than \hcop\ $1-0$ intensities (see Figure 3 of Barnes et al. 2013).
Mapping observations of \hcop\ $3-2$ and \nthp\ $3-2$ toward IRDCs with BGPS
clumps have revealed significant chemical differentiation of these two species 
on clump scales (see Battersby et al. 2010).  Our survey indicates that, while rare,
\nthp -brighter sources do occur.

Schlingman et al. (2011) explored the \nthp/\hcop\ $3-2$ ratio toward the subset of 1882 sources
observed and found a lack of correlation in the molecular intensity ratio
and 1.1 mm flux density.  The complete set of observations confirms this
lack of correlation ($r_{corr} = 0.02$; Figure 10).   
We also observe a lack of correlation in the molecular ratio 
compared to the gas kinetic temperature
measured using NH$_3$ observations ($r_{corr} = 0.01$; Figure 10).  
At $30$\as\ resolution, the
1.1 mm continuum, HHT, and GBT spectroscopic observations probe a spatial extent of 
$0.73$ pc $(D /5 \, \rm{kpc})$.  This is very similar to the median size of BGPS
clumps $0.75$ pc found by Schlingman et al.  BGPS clumps are very unlikely to be single
massive cores (median $M \sim 300$ \msun ; Schlingman et al. 2011), 
but are likely to be composed of multiple smaller cores as confirmed
by recent higher angular resolution observations (Merello et al., in prep.). 
The lack of a significant correlation
in the molecular ratio with 1.1 mm flux suggests that there is variation in the fraction
of cold, dense, CO-depleted gas between cores that are in BGPS clumps.  
This result can also be explained by the multiple
core hypothesis if the cores within a BGPS clump have different
average gas kinetic temperatures.  This hypothesis may be tested with interferometric 
observations.  The 113 sources with an intensity ratio of \nthp /\hcop $> 1$
are candidates for containing a dense, cold core within the clumps even
though the beam-averaged gas kinetic temperature spans a factor of 3.

\subsection{Linewidth and Full Width Zero Intensity}

The distributions of FWHM linewidth, $\Delta v$, 
for \hcop\ and \nthp\ $3-2$ spectra with a flag $= 1$ that are well fit by a
Gaussian line shape are shown
in Figure 11. These distributions for the full catalog are similar
to those found by Schlingman et al. (2011) with a median line width of
$3.3$ km/s for both \hcop\ $3-2$ and for \nthp\ $3-2$.  
Both distributions are positively
skewed, showing a small tail that reaches linewidths up to 16 km/s.
There are 559 \hcop\ $3-2$ detections with measured $\Delta v < 2.2$ km/s; 
some of these 559 sources may
have smaller $\Delta v$ but we are limited by the velocity resolution
(1.1 km/s) of the spectrometer.  In comparison, Wienen et al. (2012) find that
the observed linewidth of NH$_3$ (1,1) is $\sim 2$ km/s with very few sources
below $1$ km/s (their spectral resolution is a factor of two better than this
survey at $0.5$ km/s).  The \hcop\ and \nthp\ linewidths observed in this survey
are plotted against each other in Figure 11.  A few sources have \hcop\ $3-2$
linewidths that are larger than their corresponding \nthp\ $3-2$ linewidth, but
most sources cluster around the $1:1$ line.

The measured \hcop\ and \nthp\ $3-2$ linewidths are compared to the observed
1.1 flux density and gas kinetic temperatures in Figure 12.  None of these
plots are well correlated; however, there are some discernible trends.
\hcop\ $3-2$ linewidth has a lower bound that increases with 1.1 mm flux density.
For instance, sources with $S_{1.1} > 1$ Jy have 
$\Delta v$(\hcop\ $3-2) > 1.6$ km/s while sources with $S_{1.1} > 10$ Jy have 
$\Delta v$(\hcop\ $3-2) > 5$ km/s.  Similarly, there is an increasing lower
bound for linewidth with kinetic temperature.  All $\Delta v$(\hcop\ $3-2) < 1.5$ km/s
have $T_k < 20$ K while all sources with $T_k > 23$ K have 
$\Delta v$(\hcop\ $3-2) > 3$ km/s.  The highest flux density and highest gas kinetic
temperatures tend to have large ($>$ few km/s) linewidths.

The observed linewidths are typically more than an order of magnitude larger
than the expected thermal broadening in the line ($0.17$ km/s $\sqrt{T_k / 18.3 \rm{K}}$).
If the non-thermal contribution to the 
linewidths in the clumps is due to turbulent motions, then
the observed values indicate that supersonic turbulence dominates BGPS clumps.
Schlingman et al. (2011) showed that the standard supersonic turbulence
scaling relationships for molecular clouds measured in CO 
(Larson's Law; Larson 1981) de-correlates in the dense gas tracers.  This may be
due to dissipation of turbulence in the denser molecular gas; however, the observed 
linewidths are not systematically smaller than the predicted relationship for sizes
below 1 pc from Larson's Law determined from CO clouds 
(see Schlingman et al. 2011).  Alternatively,   
re-injection of turbulence by star-formation likely occurs within a subset of the 
clumps (Murray et al. 2011).  
Higher flux density sources and warmer sources are more likely to harbor embedded
protostars (see Dunham et al. 2011b) whose winds and outflows may contribute to the 
turbulent gas motions within the BGPS clump.

Assigning the entire observed linewidth to
turbulence is difficult though since unresolved bulk flows in the gas may
also contribute to $\Delta v$.  In nearby molecular clouds, velocity gradients
of $1$ km/s/kpc have been measured as well as inflowing motions along
filaments (e.g. Kirk et al. 2013).  
Examples of bulk flows may be seen in the \hcop\ $3-2$
spectra from this survey which display a self-absorbed blue asymmetry (see \S4.5).
Optical depth will also increase the linewidth.  An optical depth of
$\tau = 10$ will double the observed linewidth (Philips et al. 1979, 
Shirley et al. 2003; see \S4.4).
The observed line width is likely a combination
of unresolved bulk motions (gradients or flows), optical depth effects, 
and supersonic turbulence.
Detailed study of the kinematics within BGPS clumps requires the higher spatial
resolution currently only possible with interferometers.

We also find a lack of correlation between the \nthp\ / \hcop\ molecular ratio and the
observed linewidth.  There are BGPS source with a molecular 
intensity ratio $> 1$ and with large $> 5$ km/s linewidths.
This chemical ratio does not depend on the amount of turbulence in the clump.

The Full Width Zero Intensity (FWZI) of a spectral line is the width of the line at the 
$3 \sigma_{T_{mb}}$ level.  While the measured FWZI depends on the rms noise
level in the map, it can be determined for any line shape whereas the FWHM is 
determined from the fit of a Gaussian line shape.
We calculate the FWZI for \hcop\ $3-2$ spectra by finding the 
velocity of the first spectral channel on each side of the line peak which has a 
$T_{mb} < 3 \sigma_{T_{mb}}$.  
The resulting \hcop\ FWZI are quantized by the channel spacing ($1.1$ km/s).
For low signal-to-noise spectra, the quoted FWZI will be a lower limit to the
true value.  We only report FWZI for detections with $T_{mb} \geq 6\sigma_{T_{mb}}$
(2471 \hcop\ $3-2$ detections).

The histogram of FWZI is plotted in Figure 11.
The median FWZI for \hcop\ $3-2$ lines is 4.5 km/s.  The distribution has a tail of FWZI
out to 32.5 km/s.  The largest FWZI corresponds to source 6901 (G081.680+00.540), 
an outflow associated with the famous DR21 complex (see Davis \& Smith 1996).  
There are 38 sources with FWZI $> 15$ km/s. These objects are strong outflow candidates.

For a Gaussian line shape, the FWZI is directly related to the peak signal-to-noise
of the spectrum by
\begin{equation}
\frac{FWZI}{\Delta v} = \frac{1}{\sqrt{2}} \sqrt{\ln(T_{pk}/3\sigma_T)} \;\;\;,
\end{equation}
where $\Delta v$ is the FWHM.
The ratio of FWZI/FWHM is plotted in Figure 11 for sources with
a \hcop\ flag = $1$ and shows a 
distinct correlation that generally follows the line predicted for a Gaussian line shape.
The upward trend of FWZI/FWHM away from the dashed line in Figure 11 represent
sources with non-Gaussian line shapes indicative of asymmetries
or line wings in the spectra.  However, 
most sources in the survey have line shapes that are
consistent with a Gaussian line shape.  We discuss self-absorbed profiles in detail
in \S4.5.

\subsection{Optical Depth and Excitation Temperature}

The optical depth in the \hcop\ $3-2$ line may be determined from observations
of an isotopologue and the
radiative transfer equation assuming that the excitation temperature
of the \hcop\ and \hcopi\ $3-2$ transitions are identical 
and that there is no fractionation
between the interstellar ratio $[^{13}\rm{C}]/[^{12}\rm{C}]$ the the molecular
[\hcopi ]/[\hcop ] ratio.  The peak
optical depth of the \hcop\ $3-2$ transition is determined from the non-linear equation
\begin{equation}
\frac{T_{mb}(\rm{HCO}^+)}{T_{mb}(\rm{H}^{13}\rm{CO}^+)} = 
\frac{1 - \exp(-\tau)}{1 - \exp(-\tau\frac{[^{13}\rm{C}]}{[^{12}\rm{C}]})} \;\;.
\end{equation}
Observations were made during the final shift of observing in  
December 2012 in the \hcopi\ $3-2$ line ($260.255339$ GHz) toward
$48$ v1.0.1 sources in the Gem OB 1 star formation 
complex ($188$\degree\ $< l < 193$\degree ; see Reipurth \& Yan 2008) with 
 \hcop\ $3-2$ detections.  
Emission was detected for 34 sources ($71$\%).
Every source with a \hcopi\ $3-2$ detection has an optically thick \hcop\ $3-2$ line with  
the average optical depth for this subset of $\tau = 10.2 \pm 4.7$ assuming 
$[^{13}\rm{C}]/[^{12}\rm{C}] = 50$ (see Table 5).  Sources with \hcopi\ $3-2$ non-detections
are generally weak in \hcop\ $3-2$ resulting in an average $3 \sigma$ upper limit
of $\tau < 20$.  It is likely that even these weak \hcop\ lines are optically thick.

Due to the large optical depth observed we can solve for the excitation temperature
in the \hcop\ $3-2$ line, 
\begin{equation}
T_{ex}^{obs} = \frac{h\nu /k}{\ln\left( 1 + \frac{h\nu /k}{T_{mb}/f + J_{\nu}(T_{cmb})}\right)} \;\; ,
\end{equation}
where $f$ is the filling fraction of emission and $J_{\nu}(T) = h\nu/k / (\exp(h\nu/kT) - 1)$ is 
the Planck function in temperature units.
The observed excitation temperatures are shown in Figure 13,
assuming $f = 1$, plotted against the observed
optical depth for sources associated with Gem OB 1.  
The average excitation temperature is $T_{ex}^{obs} = 5.3 \pm 2.0$ K,
much lower than typical gas kinetic temperatures observed toward BGPS sources
with NH$_3$ $T_k$ determinations (Figure 5).
Assuming that \hcop\ emission is optically thick for all BGPS sources, we
can compare the excitation temperature with $T_k$ for this subsample.
The average $T_{ex}^{obs} = 5.5 \pm 1.4$ K is very similar to the
average found for sources in Gem OB 1.
The observed excitation temperatures are weakly correlated with the gas kinetic 
temperature (Figure 14) with sources that have $T_{ex}^{obs} > 8$ K typically
having $T_k > 20$ K.

We compare the observed $T_{ex}^{obs}$ to the excitation of the $3-2$ transition from
a single zone radiative transfer model.  The top right panel of Figure 13 shows the model
excitation temperature for single density, single kinetic temperature models
calculated using \textit{RADEX} (van der Tak et al. 2007).  The three curves correspond
to gas kinetic temperatures spanning the range of $T_k = 10$ to $30$ K.  
For the typical average volume densities
probed toward BGPS clumps of $10^3 - 10^{4.5}$ cm$^{-3}$ (Schlingman et al. 2011, Dunham et al.
2011), the model excitation temperature is $< 7$ K with $T_{ex}^{model} << T_k$.  
$T_{ex}^{model}$ 
agrees well with the observed median $T_{ex}^{obs}$ indicating that
most of the 3122 \hcop\ $3-2$ lines observed in this survey
are very sub-thermally populated.  Unfortunately, it would
a prohibitively long survey  
to determine the optical depth in all 3122 \hcop\ $3-2$ detections using \hcopi\ $3-2$
observations.

The excitation temperature must be determined or assumed to derive a column density from the observed integrated intensity.
In the optically thin, constant excitation temperature limit
($T_{ex} = T_{ex}^{CTEX} \; \forall J$), 
the total column density derived from the integrated intensity 
of \hcop\ and \nthp\ $3-2$ emission is given by
\begin{equation}
N_{\rm{thin}} = \frac{8\pi k \nu^2 f}{7 h c^3 A_{3-2}} \frac{Q(T_{ex}) J_{\nu}(T_{ex}) e^{E_u/kT_{ex}}}{J_{\nu}(T_{ex}) - J_{\nu}(T_{cmb})} \int T_{mb} \, dv \;\;,
\end{equation}
where $A_{3-2}$ is the Einstein spontaneous emission coefficent 
($1.476 \times 10^{-3}$ s$^{-1}$ for \hcop\ $3-2$ and $1.259 \times 10^{-3}$ s$^{-1}$
for \nthp\ $3-2$), $Q$ is the partition function and $E_u/k$ is the energy of the upper ($J = 3$) level 
($25.68$ K for \hcop\ $3-2$ and $26.83$ K for \nthp\ $3-2$).
For the $3-2$ transitions of both molecules, the column density becomes a sensitive
function of $T_{ex}$ below 10 K (Figure 13).  
The excitation temperature derived from Equation 3 ($T_{ex}^{obs}$) applies only for the
$J = 3$ level, but the excitation temperature for the CTEX approximation is potentially
different because of non-LTE excitation.  The bottom left panel of Figure 13
calculates $T_{ex}^{CTEX}$ from Equation 4 for a grid of models with constant density and
constant kinetic temperature assuming a total \hcop\ column density of
$3.16 \times 10^{13}$ cm$^{-2}$ ($log N = 13.5$ which is the standard assumption used
to calculate $n_{eff}$).  The plotted contours span $3.1 < T_{ex}^{CTEX} < 7$ K
for the range of average kinetic temperature measured from NH$_3$ observations.
This range overlaps the $\mean{T_{ex}^{obs}}$ calculated from Equation 3 indicating that
assuming $T_{ex}^{obs} \approx T_{ex}^{CTEX}$ in Equation 4 is a reasonable
approximation for BGPS clumps.  Using this approximation, 
the median \hcop\ column density is $2.5 \times 10^{13}$ cm$^{-2}$.

If we assume $5$ K as a typical excitation temperature
and assume the filling fraction of emission is $f \approx 1$,  
the \hcop\ $3-2$ integrated intensities may be 
converted to optically thin column densities using the scale factor 
$N_{\rm{HCO}^+}/I_{\rm{HCO}^+} = 6.9 \times 10^{12}$ cm$^{-2}$ K$^{-1}$ (km s$^{-1}$)$^{-1}$
and the \nthp\ $3-2$ emission scale factor of 
$N_{\rm{N_2H}^+}/I_{\rm{N2H}^+} = 1.0 \times 10^{13}$ cm$^{-2}$ K$^{-1}$ (km s$^{-1}$)$^{-1}$.
The corresponding $3\sigma$ upper limit to the optically thin 
\hcop\ $3-2$ column density in this survey 
is $N_{3\sigma} > 1.2 \times 10^{12}$ cm$^{-2}$.
However, we know that the \hcop\ $3-2$ line is likely optically thick from
the results of \hcopi\ $3-2$ observations;  therefore, we should apply a correction
to the optical depth of $N = N_{thin} \tau / (1 - e^{\tau})$ (Goldsmith \& Langer 1999).
The \hcop\ column densities shown in Figure 14 are likely lower limits because we do not
know the true optical depth for sources outside Gem OB 1.

We plot the \hcop\ column density versus the total H$_2$ column
density calculated from the 1.1 mm peak flux density assuming Ossenkopf \& Henning
(OH5) opacities and that
the dust temperature is equal to the gas kinetic temperature from NH$_3$
observations (Figure 14).  Gas kinetic temperature and dust temperature
are only expected to be well coupled at high densities ($n > 10^5$ cm$^{-3}$;
see Goldsmith 2000); however, in the absence of other information, it is a
better approximation than assuming all sources are at the same dust temperature.
The plotted column density errorbars are purely statistical errors based on the
uncertainty in $\sigma_{T_{mb}}$.  The total uncertainty in
the \hcop\ column density is dominated by the uncertainty in the unknown filling fraction
of emission and the uncertainty in the unknown optical 
depth which would likely increase the column density by factors of a few.
With these caveats in mind, we plot the \hcop\ column density and abundances
in Figure 14.  In contrast with the integrated intensity correlation (\S4.2),
there is no correlation between the \hcop\ and H$_2$ column densities. 
This result again highlights the importance of the excitation conditions
and not the total molecular column density in driving the observed intensity
correlations.  A similar lack of correlation was also observed for NH$_3$ column densities
versus H$_2$ column densities by Wienen et al. (2012).
We find a median \hcop\ abundance of $1.6 \times 10^{-9}$.  
The \hcop\ abundance also shows no discernible
trend with Galactocentric radius (Figure 14).  This result is in contrast with NH$_3$
observations which show a decrease by a factor of 7 in abundance with 
Galactocentric radius (Dunham et al. 2011a).  Both studies are limited by a paucity of 
sources at large Galactocentric radius with measured gas kinetic temperatures.

Unfortunately, we cannot easily calculate the optical depth in the \nthp\ $3-2$
line because the 45 hyperfine lines are too heavily blended to obtain 
believable constraints.  Furthermore, observations of the $^{15}\rm{NNH}^+$ isotopologue
is prohibitive because the interstellar $[^{15}\rm{N}]/[^{14}\rm{N}]$ abundance is a factor
of $5 - 10$ smaller than the ISM $[^{13}\rm{C}]/[^{12}\rm{C}]$ abundance (Wilson \& Rood 1994, 
Adande \& Ziurys 2012). 
\nthp\ $3-2$ is also very likely sub-thermally populated based on our radiative
transfer calculations.  Given the more than order of magnitude uncertainties,
we do not report column densities for \nthp\ $3-2$ emission.
We recommend observations of the \nthp\ $1-0$ line where the hyperfine splitting is
more easily resolved (e.g., Pirogov et al. 2003, Pirogov et al. 2007, Reiter et al. 2011a, Barnes et al. 2013).

\subsection{Self-absorbed Line Profiles}

While most \hcop\ $3-2$ lines in the survey are believed to have $\tau > 1$
based on limited H$^{13}$CO$^+$ $3-2$ observations and comparison to 
radiative transfer models (\S4.4),
there is a small percentage of sources that display clear line asymmetries 
indicative of kinematic motions coupled with very optically thick lines.
For a \hcop\ $3-2$ line profile to be classified as self-absorbed, the
profile must show two peaks and an absorption dip over the span
of at least 3 channels ($3.3$ km/s) with the \nthp\ $3-2$ line 
profile having a single-peak (\hcop\ flag = $3$ while \nthp\ flag = $1$).  
This situation occurs for $80$ ($1.3$\%) sources (Table 5).
Self-absorbed line profiles are extremely rare in this survey.  This may,
in part, be due to the low spectral resolution ($1.1$ km/s) 
required to cover the full velocity range ($-60$ to $170$ km/s) 
of sources in the longitude range observed in the first quadrant.

We may classify the self-absorbed sources sources as blue, red, or equal asymmetries by
the location of the $v_{lsr}$ maximum.  $47$ sources have a clear
blue asymmetry while $26$ sources have a clear red asymmetry and $7$
sources have equal height peaks.  The blue excess for the subset of sources 
showing distinct self-absorption is $(N_B - N_R)/N_{SA} = 0.57$.  
These $80$ sources (listed in Table 4) are excellent high-mass, large-scale collapse 
candidates (see Reiter et al. 2011b) and should be followed up at higher 
spatial resolution.

\section{Conclusions}

In this paper, we have presented a complete spectroscopic catalog for all
sources in the BGPS v1.0.1 source catalog with $7.5$\degree $\leq l \leq 194$\degree\
and characterized the properties of the molecular emission. 3126 sources ($50.5$\%)
were detected with unique $v_{LSR}$.  
\hcop\ $3-2$ is a significantly better kinematic tracer
of BGPS clumps than CO isotopologues with
contamination from multiple velocity components rare (only $1.3$\%).
The detection fraction of BGPS v1.0.1 sources is $> 50$\%\ for 1.1 mm sources
with flux densities $> 200$ mJy and $> 90$\%\ for flux densities $> 500$ mJy.
The majority of BGPS clumps appear to be physically associated with other clumps.
This is the largest targeted dense gas survey
of the Milky Way to date and traces the major spiral arm structures of the Milky Way
in Galactocentric radius.

\hcop\ and \nthp\ molecular intensities are well correlated with 1.1 mm flux density and with 
each other while molecular column densities are not well correlated with the total
H$_2$ column density.  \hcop\ $3-2$ intensity is also correlated with gas kinetic temperature.
\hcop\ $3-2$ emission is likely optically thick and sub-thermally
populated ($T_{ex} \sim 5$ K $<< T_k$) toward most BPGS clumps.
These observed intensity correlations are most likely due to the sensitivity of the $3-2$ transitions
to excitation conditions probed by the dense gas.  The molecular intensity ratio does not
correlate with 1.1 mm flux density, $T_k$, or $\Delta v$.  BGPS clumps are likely
composed of smaller cores which may have a range of different chemical and excitation 
conditions.  Sources with large \nthp /\hcop\ intensity ratios probably contain
significant reservoirs of cold, dense gas.  The median $\Delta v = 3.3$ km/s (FWHM) and is
consistent with supersonic turbulence although optical depth and bulk motions
within the core are contributors to the linewidth.  The observed linewidth
does not correlate with 1.1 mm flux density or $T_k$ although the lower bound
of $\Delta v$ does correlated with both quantities.

In the coming months, source catalogs from the ATLASGAL and Hi-Gal continuum
surveys of the Galaxy will be released.  The kinematic data provided
in this paper provides the necessary information for calculating
kinematic distances.  Future work by the BGPS team will use this spectroscopic catalog
and expand the priors to better
constrain Distance Probability Density Functions and statistically
analyze the physical properties of BGPS clumps in different evolutionary stages
and in different Galactic environments.



\section*{Acknowledgments}

We sincerely thank the staff of the Arizona Radio Observatory - 
in particular, the operators Bob Moulton, John Downey, 
Patrick Fimbres, and Craig Sinclair - 
for their help and hospitality during observing. We also sincerely thank
the referee for comments that improved the manuscript.
This work was supported by NSF grant AST-1008577.





\begin{figure}
\figurenum{1}
\epsscale{1.0}
\plotone{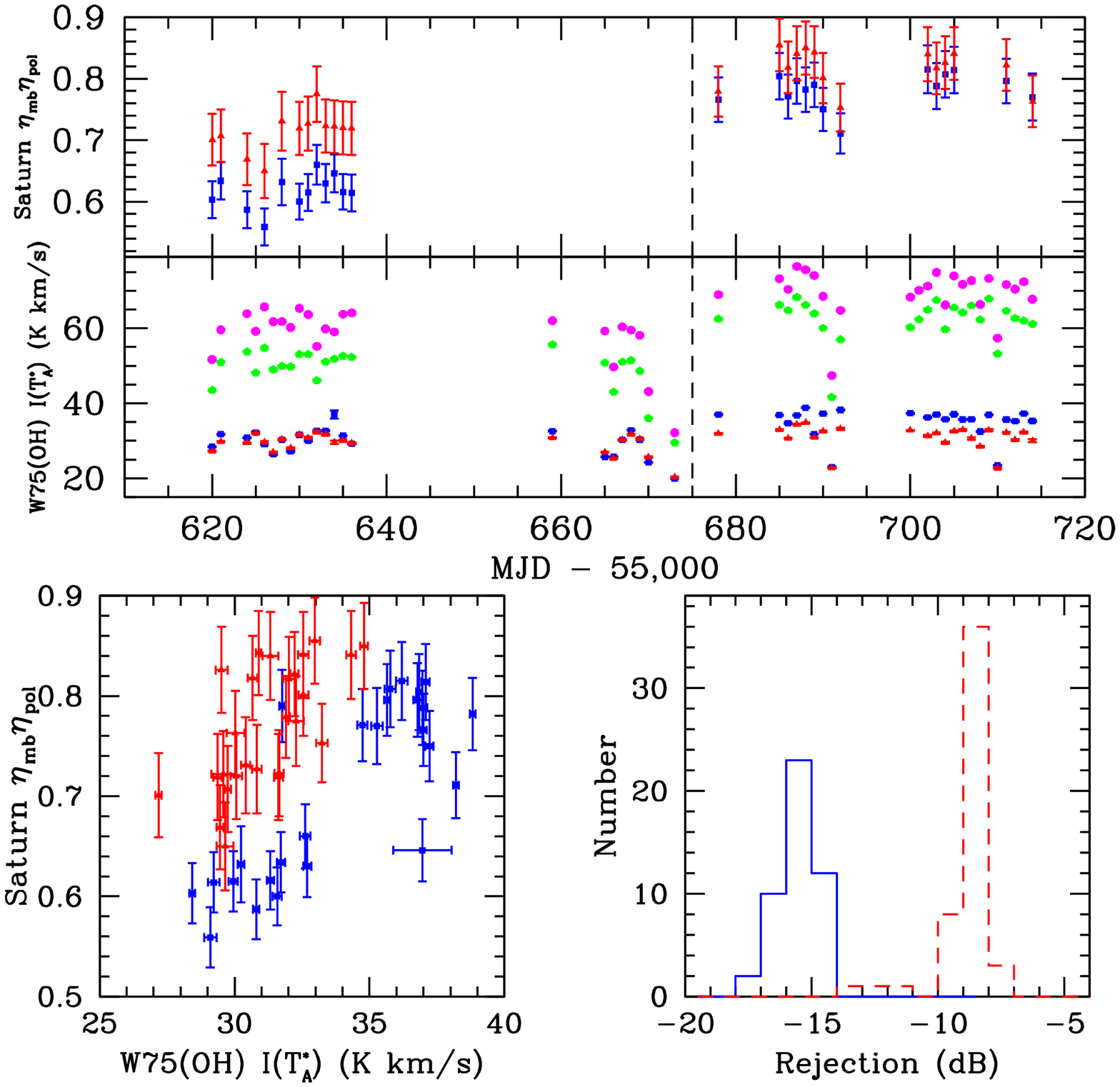}
\figcaption{TOP: The main beam efficiency of the USB measured on Saturn  
and the integrated intensity of \hcop\ $3-2$ (LSB) and \nthp\ $3-2$ (USB) emission observed
toward W75(OH) for Hpol USB (blue squares), Vpol USB (red triangles), Hpol LSB (green pentagon),
and Vpol LSB (magenta circle).  Due to PH$_3$ absorption in Saturn's
atmosphere in the LSB, we only plot USB data for Saturn.  The dashed line MJD-55,000 = 675 corresponds to a systematic
change in the calibration associated with a warmup and adjustment of the receiver optics.
BOTTOM LEFT:  The USB main beam efficiency of Saturn is well correlated with integrated intensities
measured for W75(OH).  
BOTTOM RIGHT: Histograms of the sideband rejection in the USB for
Hpol (blue solid line) and Vpol (red dashed line).}
\end{figure}


\begin{figure}
\figurenum{2}
\epsscale{1.0}
\plotone{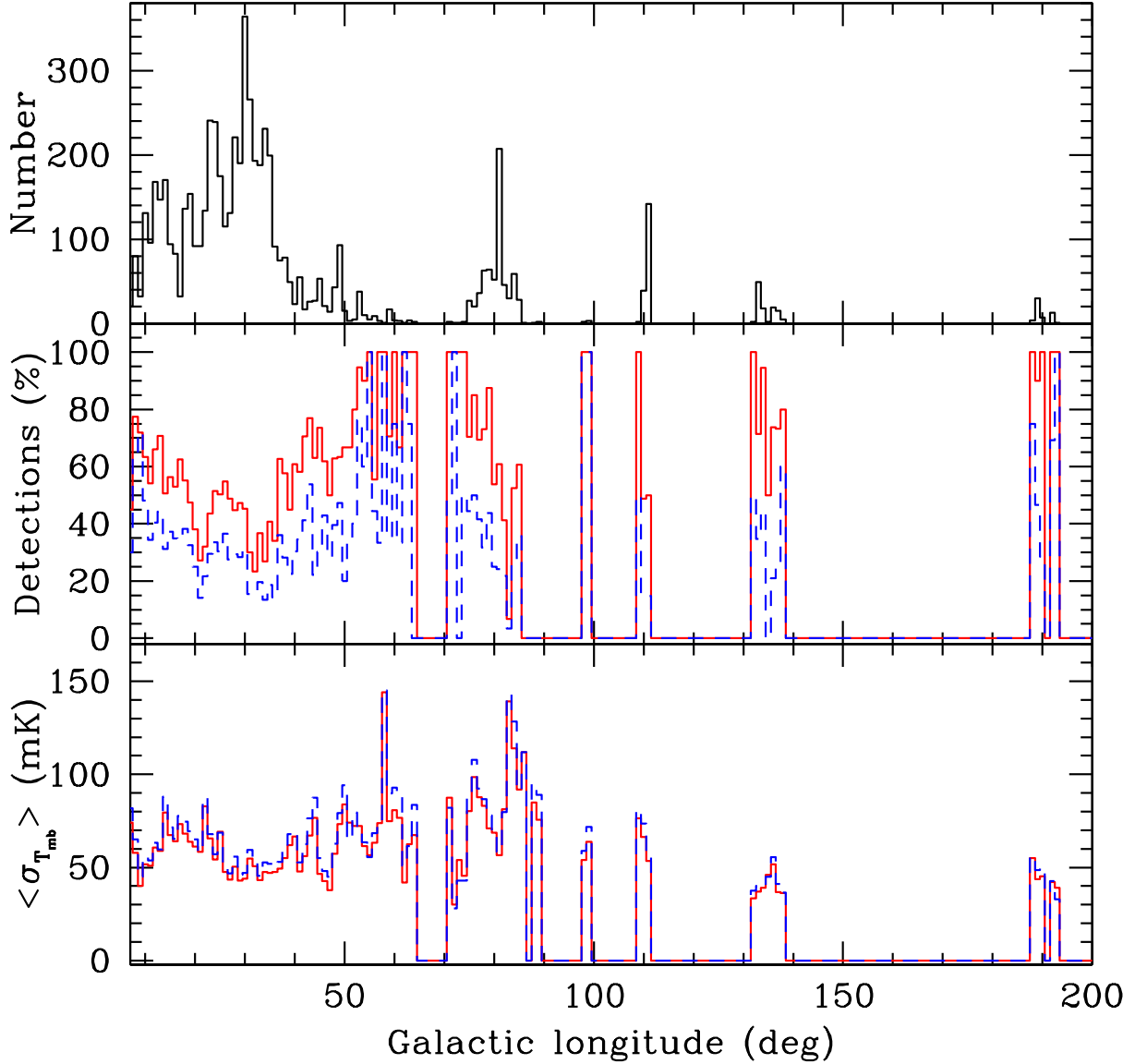}
\figcaption{TOP: The distribution of BGPS v1.0.1 catalog sources with
Galactic longitude.
MIDDLE: Detection fraction for each Galactic longitude bin of \hcop\ $3-2$
(solid red histogram) and \nthp\ $3-2$ detections
(dashed blue histogram).  BOTTOM:  The distribution of the
average baseline rms for \hcop\ $3-2$ (red, solid line) and \nthp\ $3-2$ (blue, dashed line).
Values of zero indicate l ranges with no sources.  $\mean{\sigma_{T_{mb}}} = 58$ mK for
\hcop\ $3-2$ and $\mean{\sigma_{T_{mb}}} = 62$ mK for
\nthp\ $3-2$ for the complete spectroscopic survey.}
\end{figure}


\begin{figure}
\figurenum{3}
\plottwo{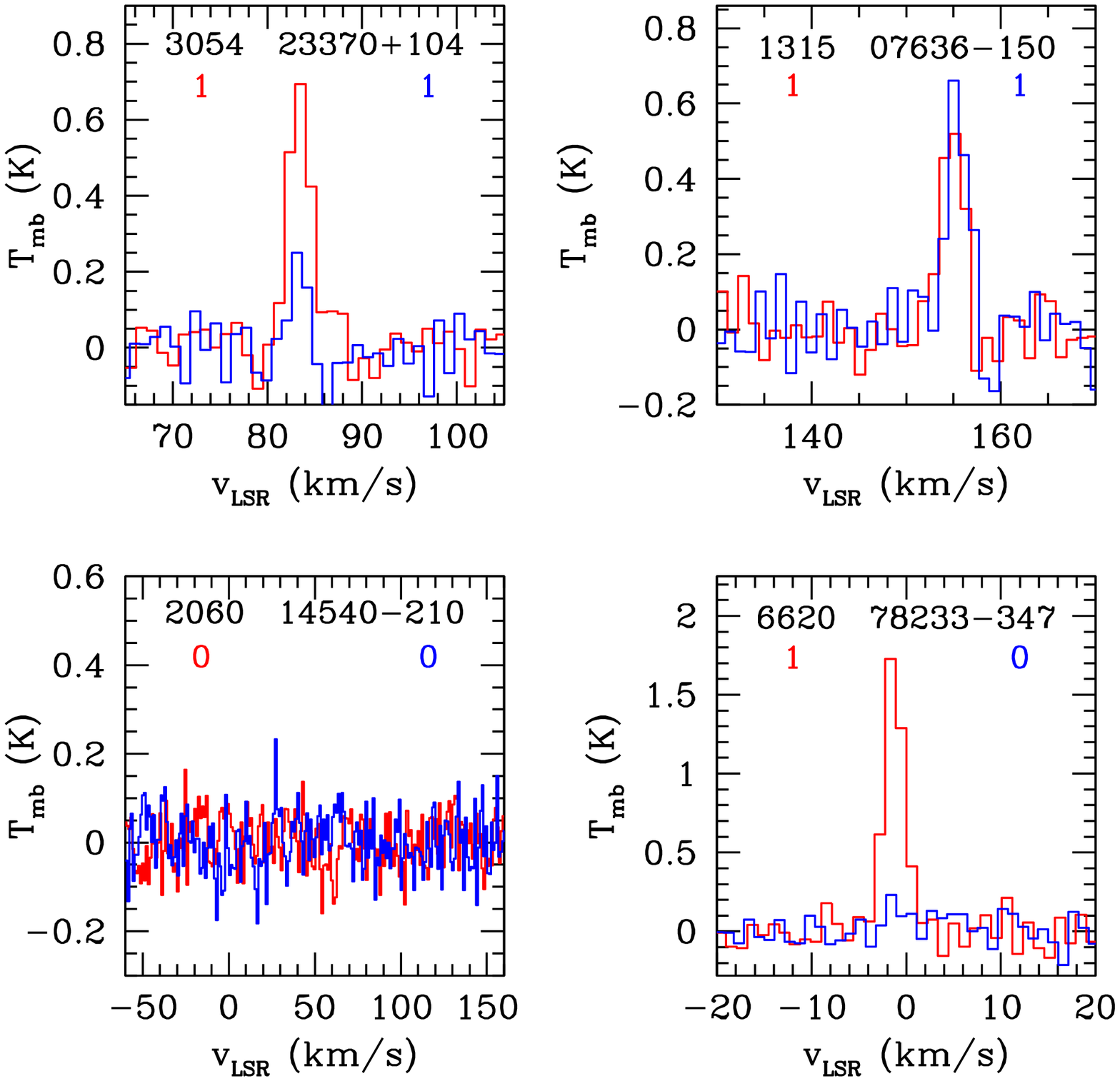}{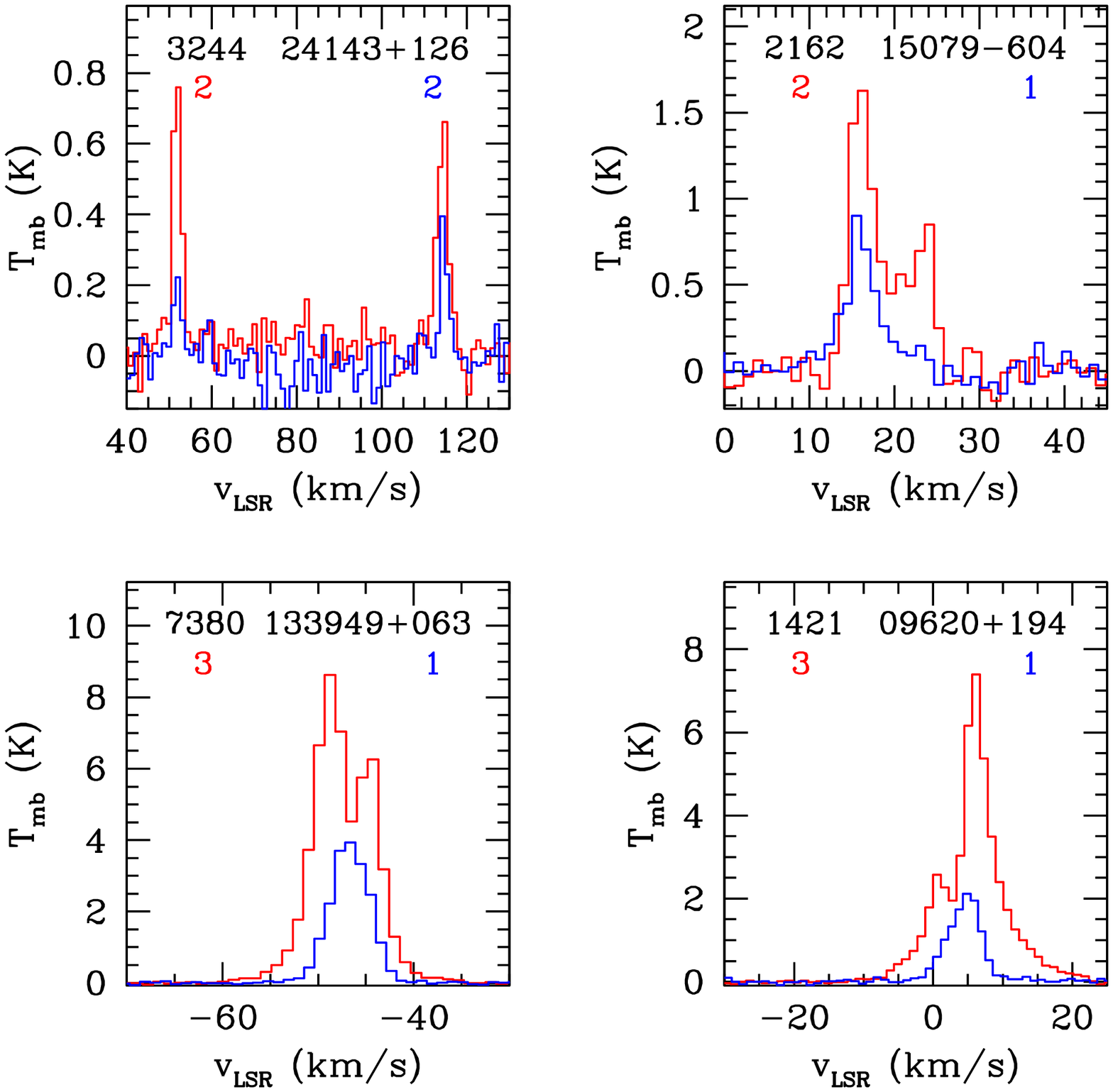}
\figcaption{Example spectra showing different flag combinations.  Red spectra are
\hcop\ $3-2$ and blue spectra are \nthp\ $3-2$.  The Bolocat v1.0.1 source number is listed
in the upper left of each panel while the truncated 
v1.0.1 catalog source name is listed in the top right of
each panel.  The \hcop\ $3-2$ flag is shown in red below the source number 
on the left while the \nthp\ $3-2$ flag is
shown in blue below the source name on the right.  The majority of detections are similar
to source 3054 (G23.370+0.104).
Source 1315 (G07.636-0.150) is an example of a \nthp\ brighter source.
Source 6620 (G78.233-0.347) is an example of a \hcop\ detection and a \nthp\ non detection.
Source 2060 (G14.540-0.210) is a non detection in both lines.
Note that the velocity scale for the non-detections source 2060
spans the full range of plausible $v_{LSR}$ for a source at $l = 14.5$\degree .  The spike
near 30 km/s in the G14.540-0.210 \nthp\ spectrum is not a detection 
as it is only one channel wide
and does not appear independently in both polarizations.  
Sources 3244 (G24.143+0.126) and 2162 (G15.079-0.604) 
show examples of multiple $v_{LSR}$ components where \nthp\ peaks
in one or both velocity components.
The two self-absorbed profiles (sources 7380 and 1421) show examples of blue and red asymmetries in the
\hcop\ $3-2$ profiles respectively with the \nthp\ $3-2$ spectra peaking near the self-absorption
minima.}
\end{figure}


\begin{figure}
\figurenum{4}
\epsscale{1.0}
  \centering
   \vspace*{10.5cm}
   \leavevmode
   \includegraphics{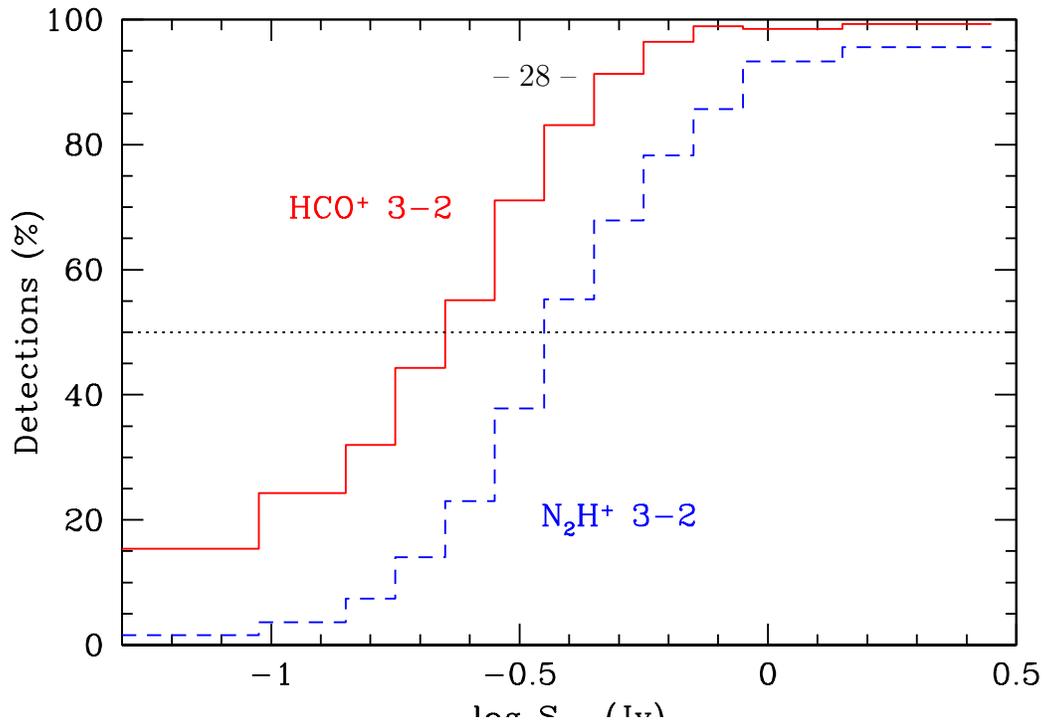}
   \includegraphics{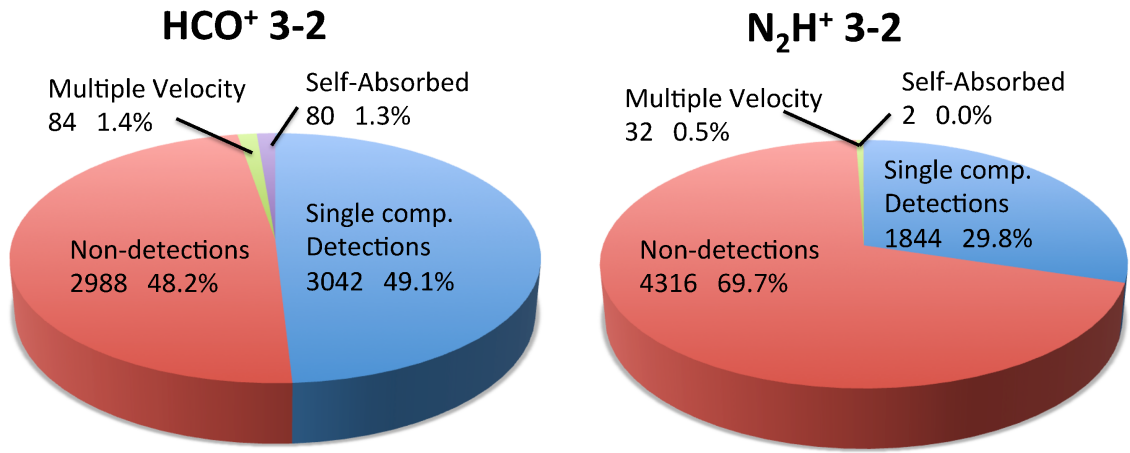}

 \vskip 0.2in
\parbox{6.3in}{\caption{TOP: The detection fraction for \hcop\ $3-2$ (red solid line)
and \nthp\ $3-2$ (blue dashed line) versus 1.1 mm flux densities derived from the v2.0 BGPS
maps at the location of the HHT pointing position. The \hcop\ $3-2$ detection fraction 
is above 50\% for 1.1 mm flux densities $ > 200$ mJy.  BOTTOM: Detection flag statistics for
\hcop\ and \nthp\ $3-2$ for 6194 sources in the complete spectroscopic catalog.
}}

\end{figure}


\begin{figure}
\figurenum{5}
\epsscale{1.0}
\plotone{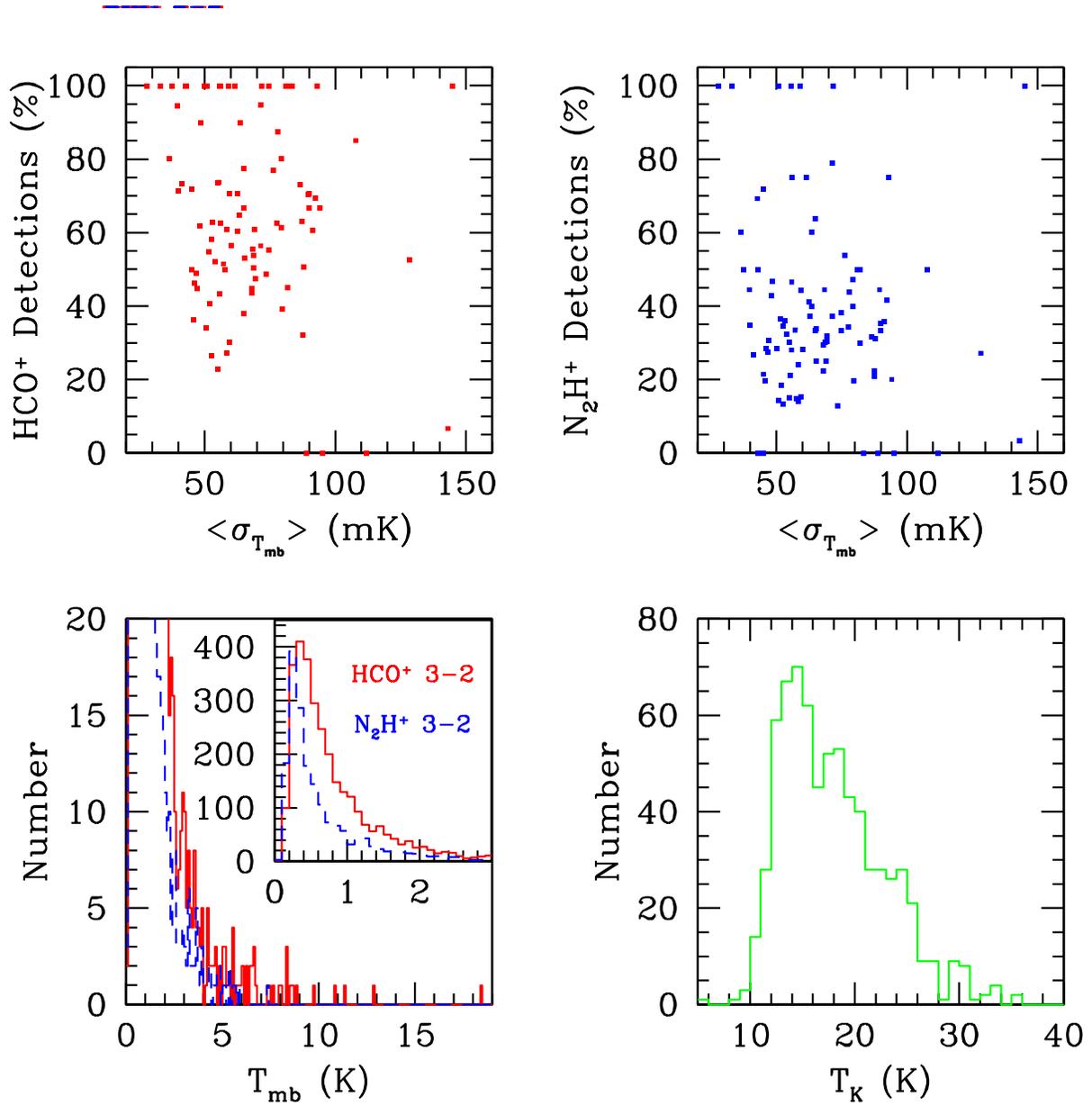}
\figcaption{TOP PANELS: The \hcop\ $3-2$ (LEFT) and \nthp\ $3-2$ (RIGHT) detection fraction plotted
versus average baseline rms per $1$\degree\ bin in Galactic longitude.
BOTTOM LEFT: The main beam brightness temperature for \hcop\ $3-2$ (red solid line) and \nthp\ $3-2$ (blue dashed line).  We have truncated the histograms at 3K to better display the
peak of each distribution. BOTTOM RIGHT: The histogram of gas kinetic temperature from the overlap NH$_3$ sample. }
\end{figure}


\begin{figure}
\figurenum{6}
\epsscale{1.0}
   \centering
   \vspace*{9.5cm}
   \leavevmode
   \includegraphics{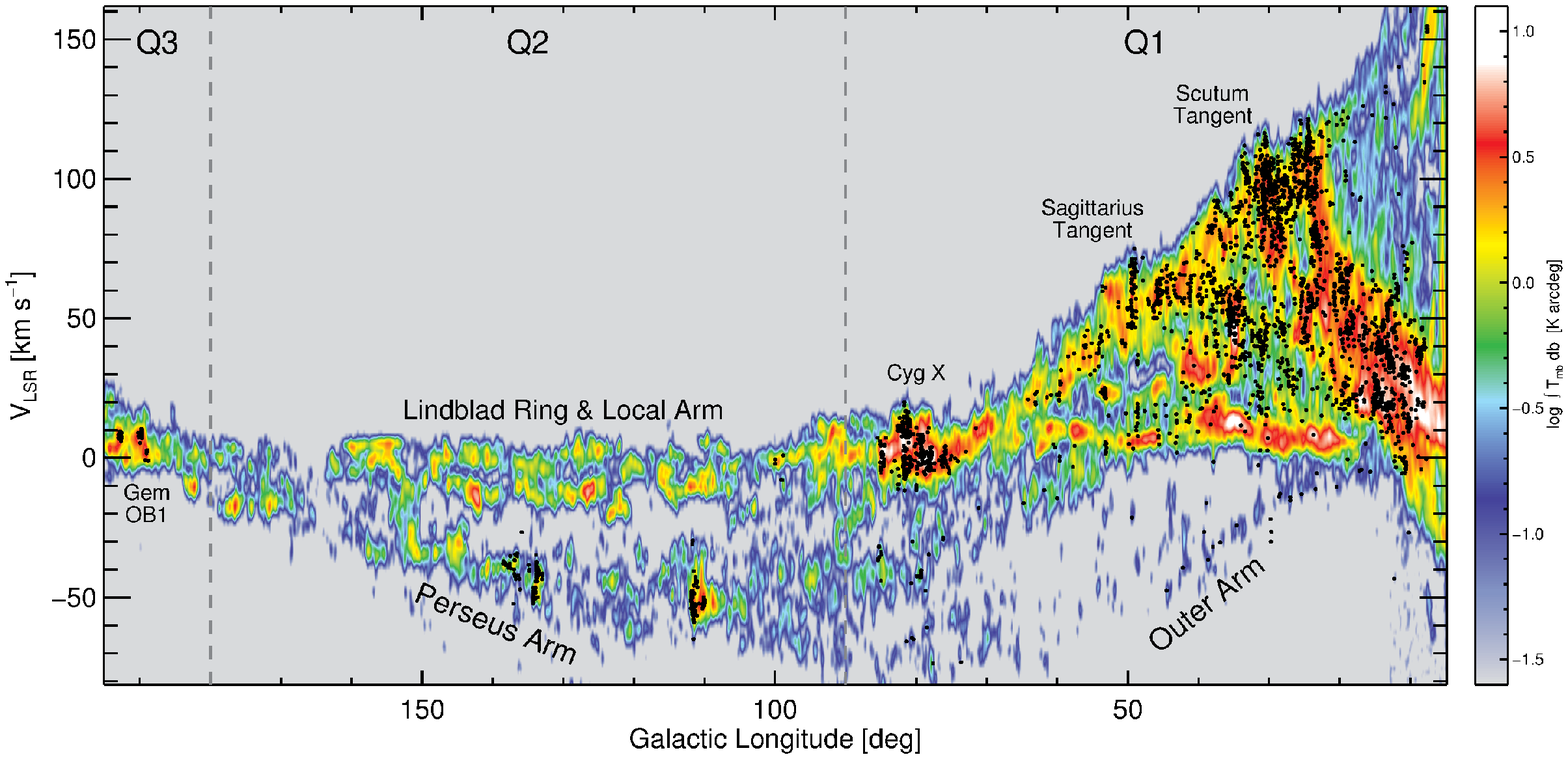}
   \includegraphics{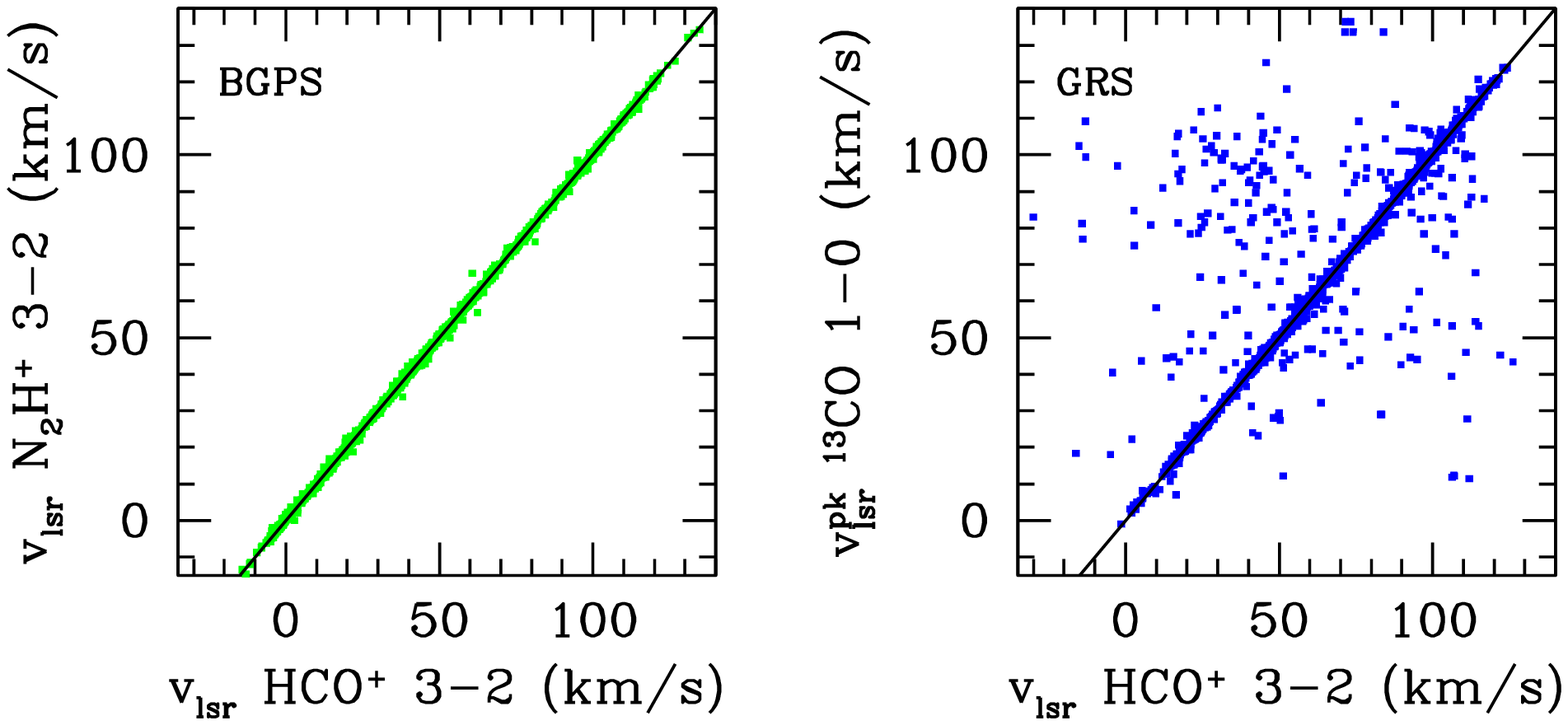}

 \vskip 0.2in
\parbox{6.3in}{\caption{TOP: Velocities of BGPS clumps (black circles) 
overlaid on the CO $1-0$ l-v diagram of the Milky Way from Dame et al. (2001).  BOTTOM PANELS: (LEFT) $v_{lsr}$ of \nthp\ $3-2$ vs. \hcop\ $3-2$ for 1611 BGPS clumps with single component detections in both lines.  The black line is the $1:1$ line.  (RIGHT) $v_{lsr}$ of $^{13}$CO $1-0$ vs. \hcop\ $3-2$
for 1681 GRS sources that overlap BGPS molecular detections. 
}}

\end{figure}


\begin{figure}
\figurenum{7}
\epsscale{0.9}
\plotone{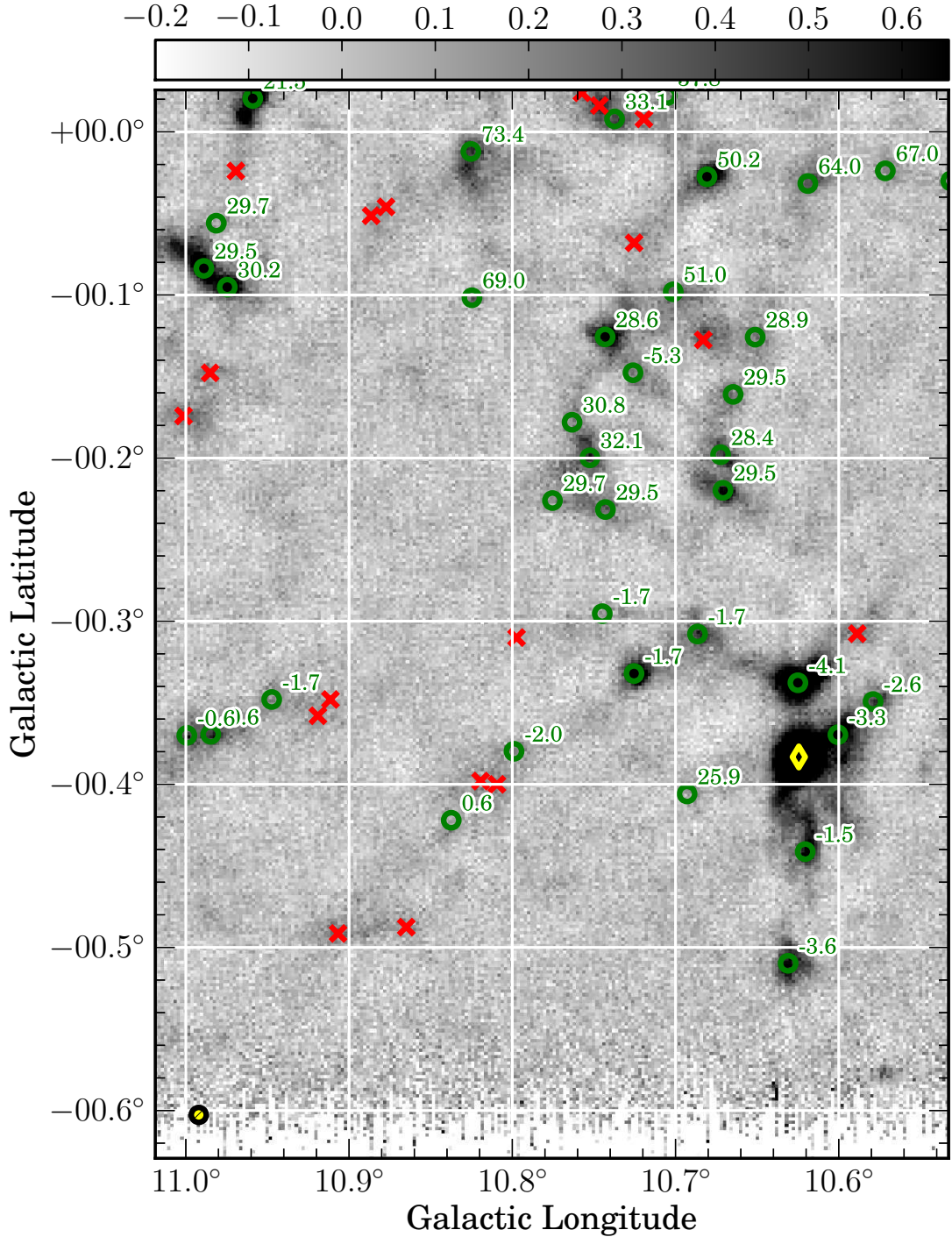}
\figcaption{Example finder chart for the $0.5$\degree\ region
centered on $l = 10.5$\degree\ $b = -0.3$\degree .  
Velocities of BGPS clumps (km/s) overlaid on the BGPS 1.1 mm v2.0
continuum image. Green circles indicate a single \vlsr\ component detection, 
yellow diamonds indicate multiple
\vlsr\ detected, and rojo crosses indicate no \hcop\ or \nthp\ $3-2$ detection.  The grey scale
is flux density in units of Jy. The Bolocam beam at 1.1 mm is shown as the filled yellow circle
in the lower left.}
\end{figure}


\begin{figure}
\figurenum{8}
\epsscale{1.0}
\plotone{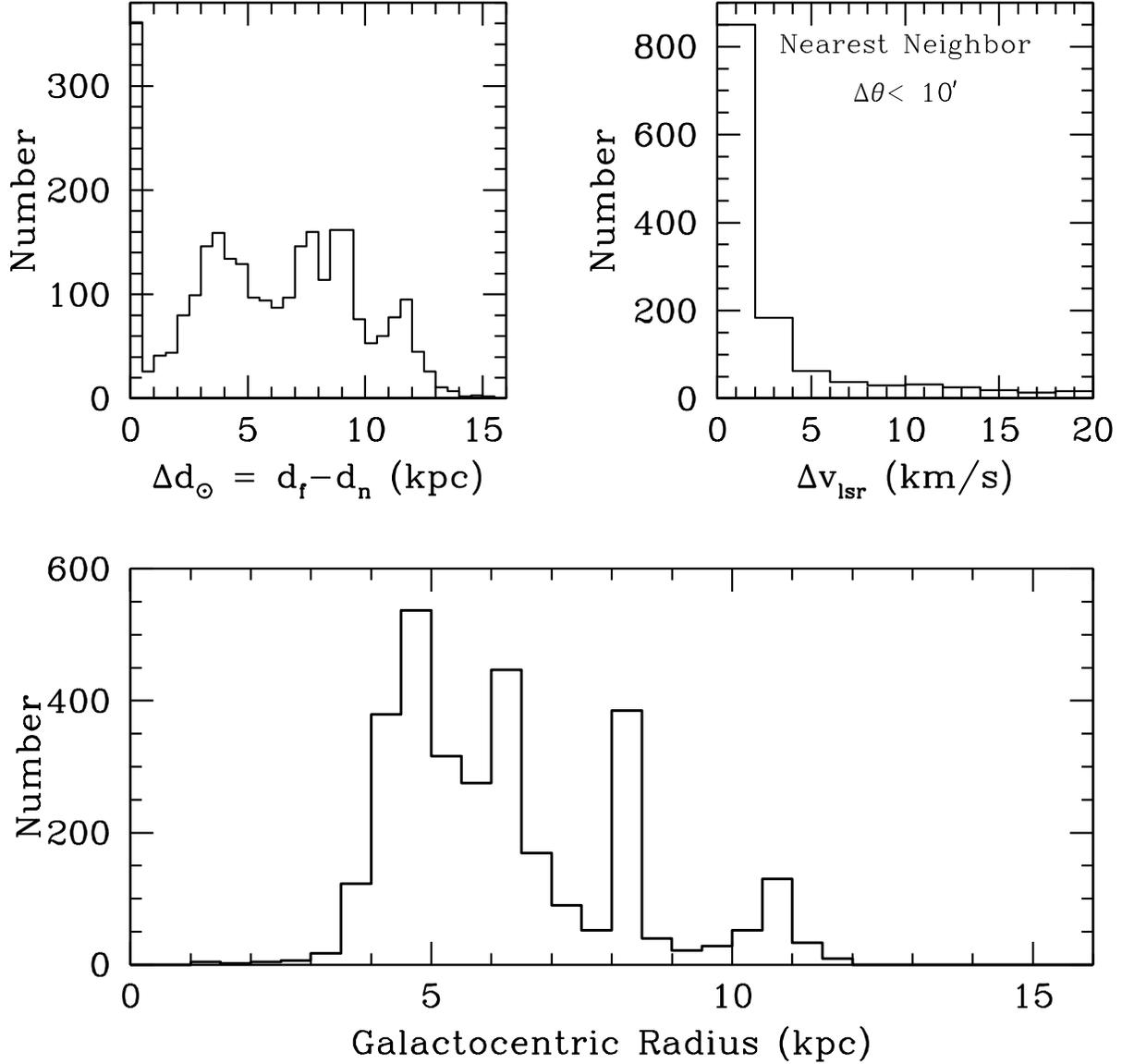}
\figcaption{TOP LEFT: The distribution of the difference between the far and near kinematic distances.  TOP RIGHT: The distributions of nearest neighbor sources in $\Delta v_{LSR}$ for neighbors that are within $10$\arcmin .  BOTTOM: The distribution of sources in Galactocentric radius.  Peaks are clearly distinguished for the molecular ring, Sagittarius arm, local spur, and Perseus arm.}
\end{figure}


\begin{figure}
\figurenum{9}
\epsscale{1.0}
\plotone{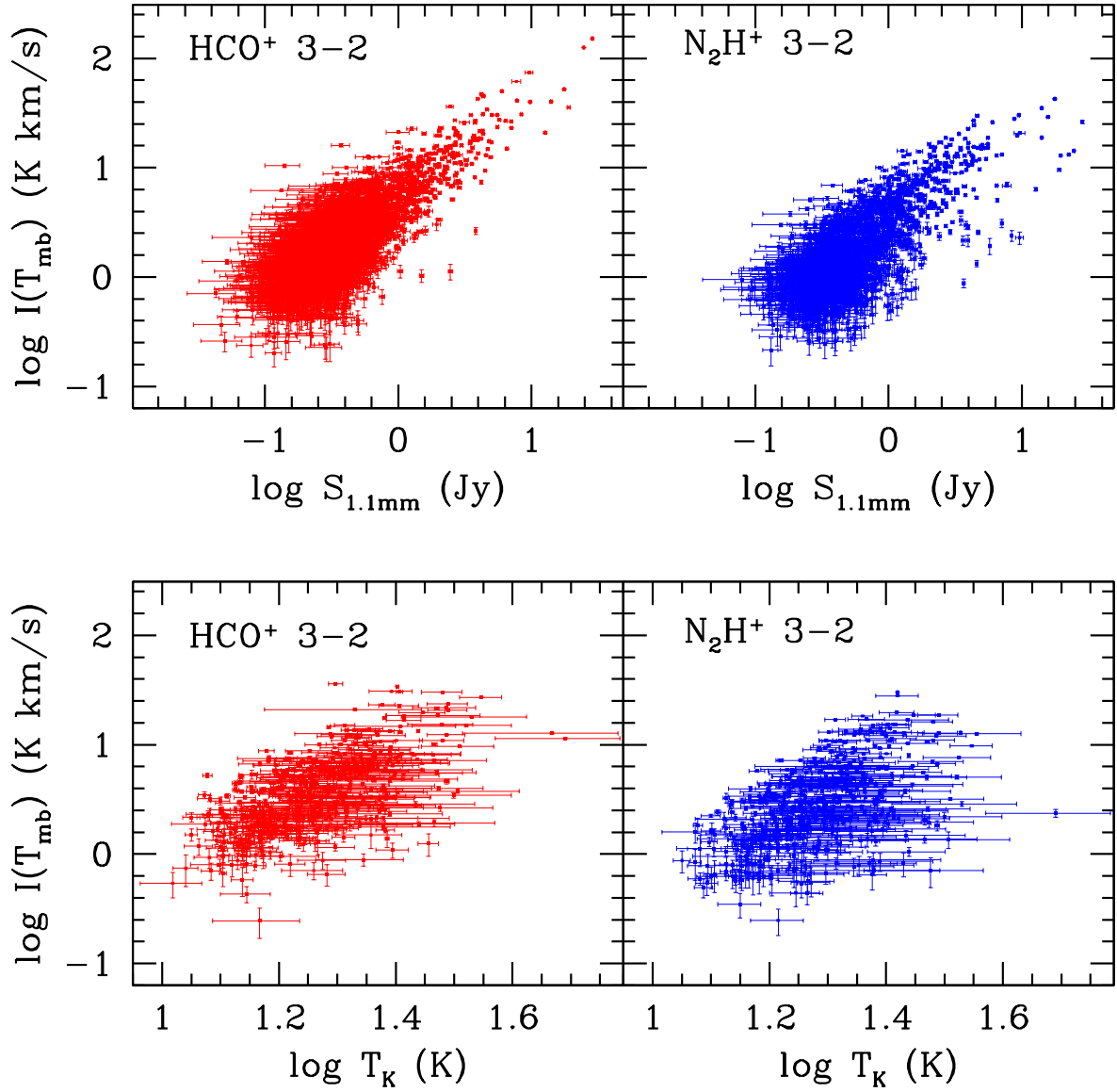}
\figcaption{The integrated intensity of
\hcop\ $3-2$ (top left) and \nthp\ $3-2$ (top right) 
vs. $1.1$ mm flux density from v2.0 maps.   The integrated intensity of
\hcop\ $3-2$ (bottom left) and \nthp\ $3-2$ (bottom right) 
vs. gas kinetic temperature determined from NH$_3$ observations.}
\end{figure}


\begin{figure}
\figurenum{10}
\epsscale{1.0}
\plotone{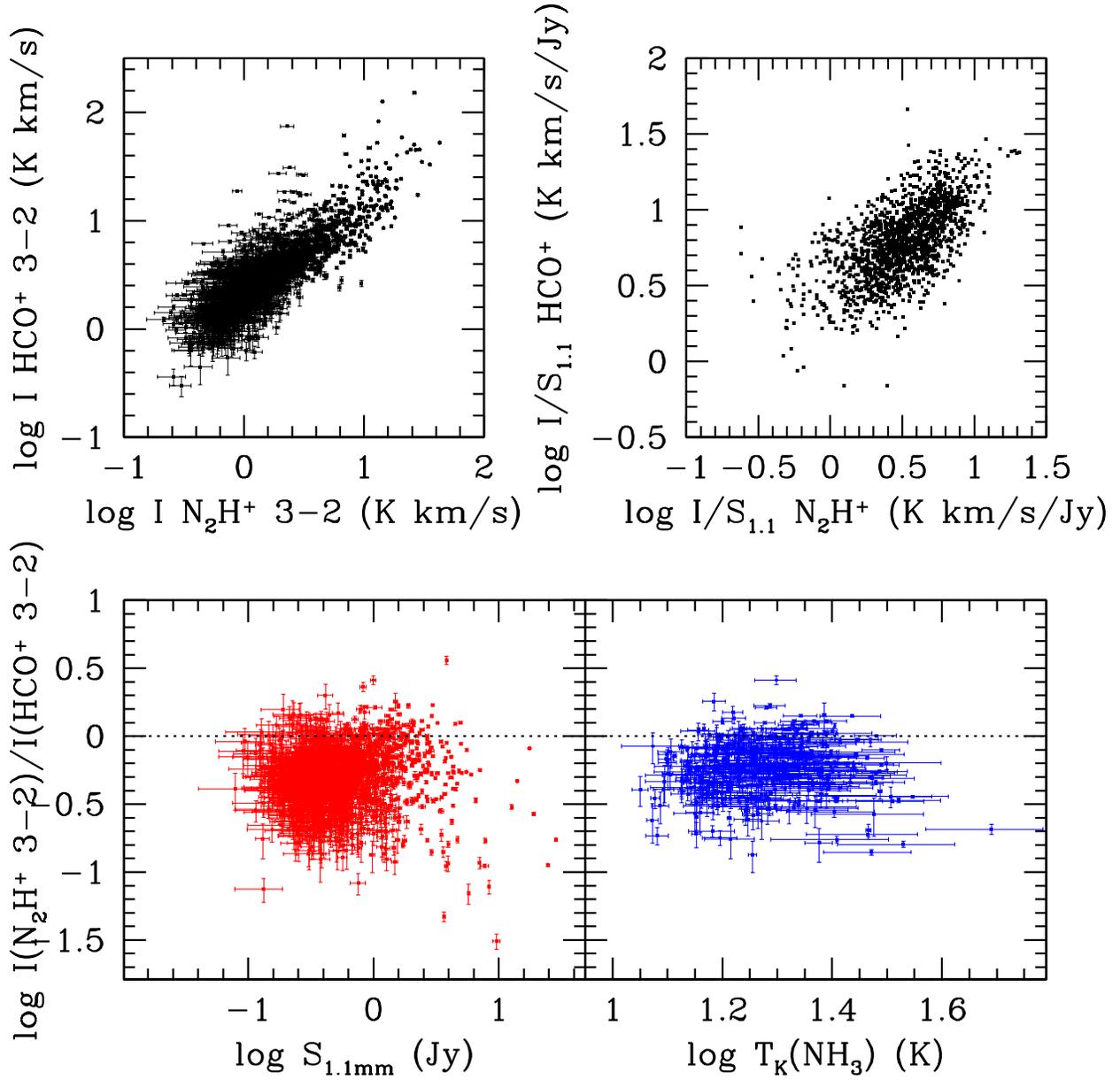}
\figcaption{TOP LEFT: A strong correlation 
is observed between the \hcop\ and \nthp\ $3-2$ integrated intensities.
TOP RIGHT: A weaker correlation is observed between \hcop\ and \nthp\ $3-2$ 
integrated intensity to 1.1 mm flux density ratio.  Errorbars have been suppressed 
to better display the points.
BOTTOM LEFT: The integrated intensity ratio
\nthp\ / \hcop\ $3-2$  vs. $1.1$ mm flux density from v2.0 maps.   
BOTTOM RIGHT: The integrated intensity ratio \nthp\ / \hcop\ $3-2$ 
vs. gas kinetic temperature determined from NH$_3$ observations.
}
\end{figure}


\begin{figure}
\figurenum{11}
\epsscale{0.9}
\plotone{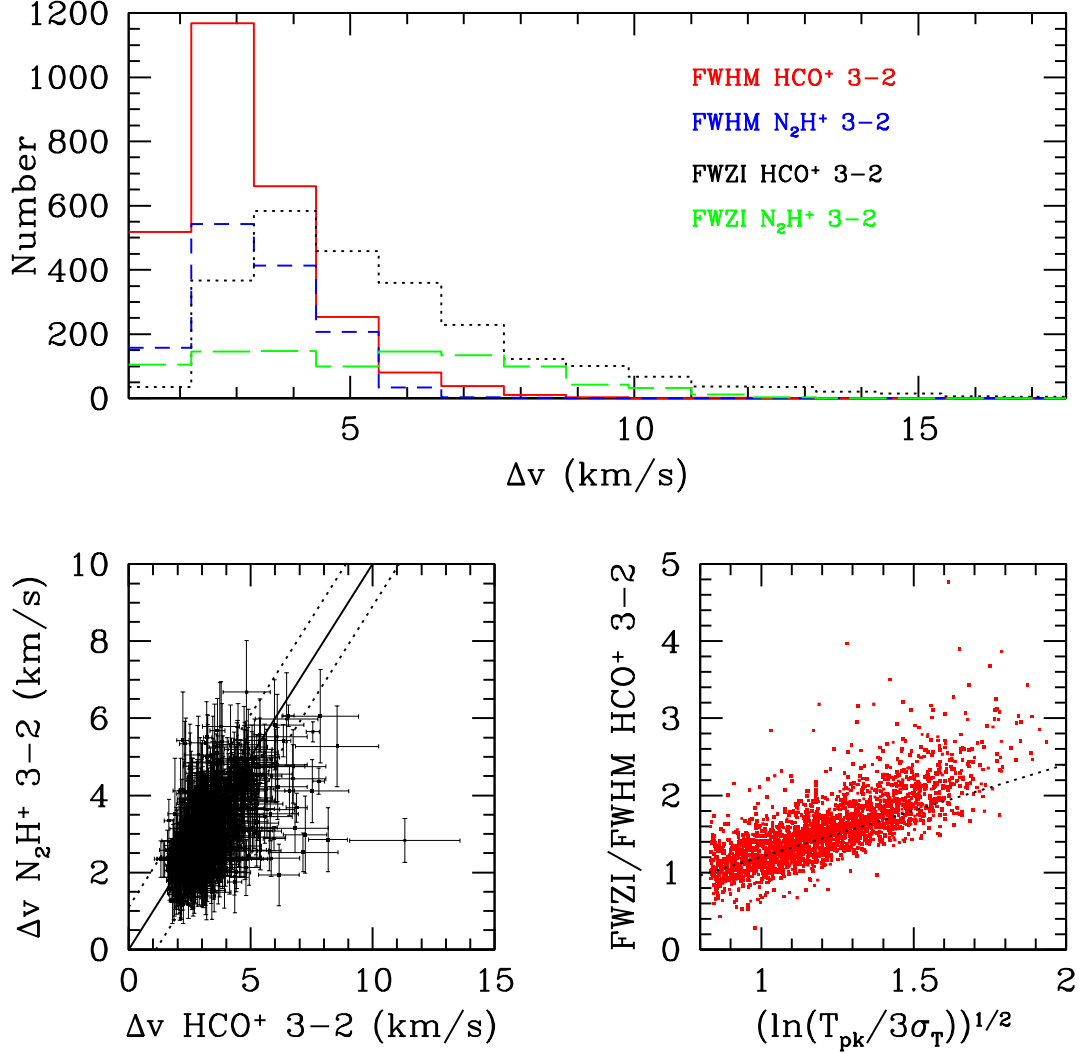}
\figcaption{TOP: Histograms of \hcop\ $3-2$ FWHM (solid red line), \nthp\ $3-2$
FWHM (short-dashed blue line), \hcop\ $3-2$ FWZI (doted black line), and \nthp\ $3-2$ FWZI (long-dashed
green line).  BOTTOM LEFT: A correlation is observed between the \nthp\ $3-2$ and \hcop\ $3-2$
linewidths.  A solid 1:1 line is plotted with dashed lines offset by the velocity resolution
of one channel (1.1 km/s).
BOTTOM RIGHT: The ratio of FWZI/FWHM for \hcop\ $3-2$ correlates against $\sqrt{\ln(T_{pk}/3\sigma_{T_{mb}})}$.  The dashed line shows the expected relationship for a Gaussian line shape.
}
\end{figure}


\begin{figure}
\figurenum{12}
\epsscale{0.9}
\plotone{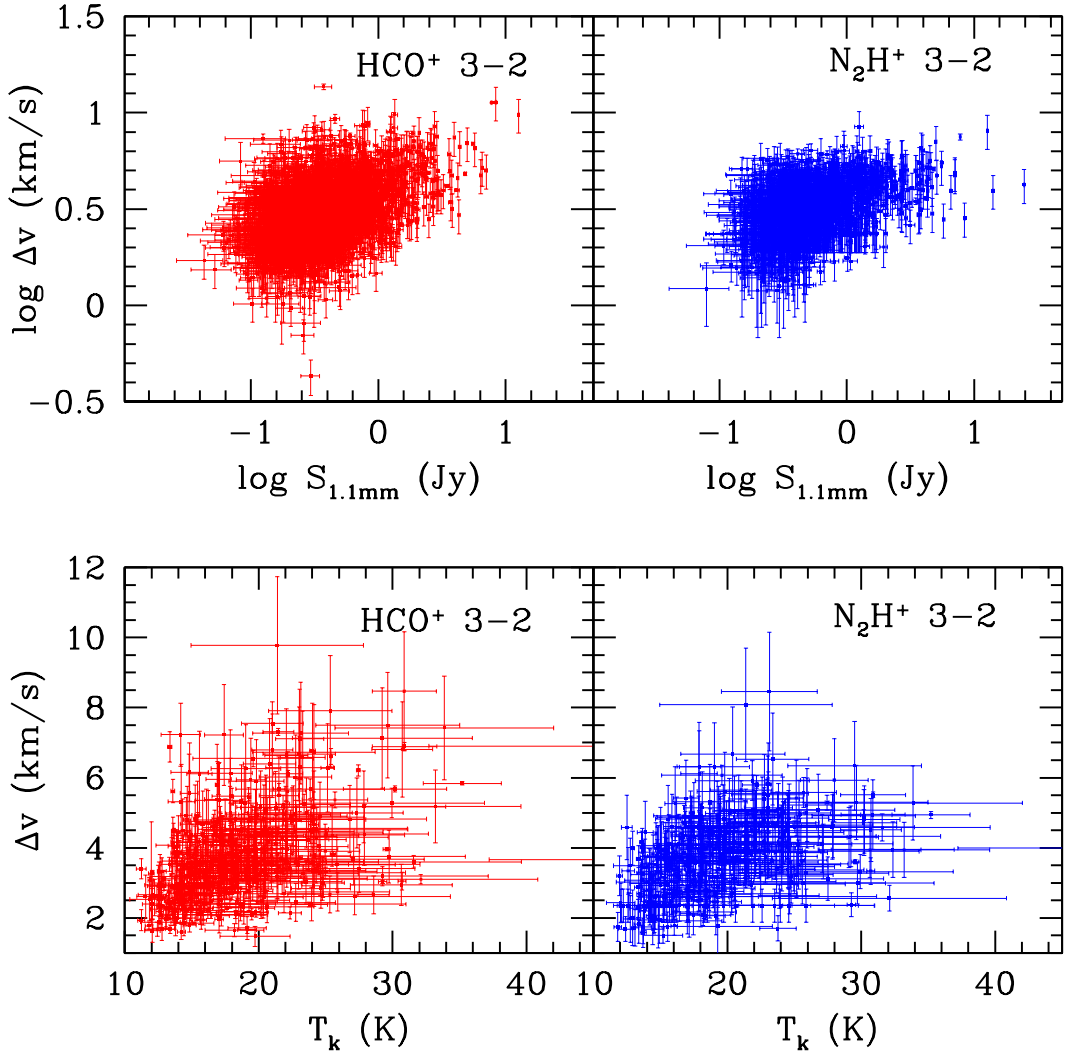}
\figcaption{The \hcop\ $3-2$ FWHM (top left) and the \nthp\ $3-2$ FWHM (top right) 
vs. 1.1 mm flux density.  The \hcop\ $3-2$
FWHM (bottom left) and the \nthp\ $3-2$ (bottom right) vs. gas kinetic temperature. }
\end{figure}


\begin{figure}
\figurenum{13}
\vspace*{10.5cm}
\includegraphics{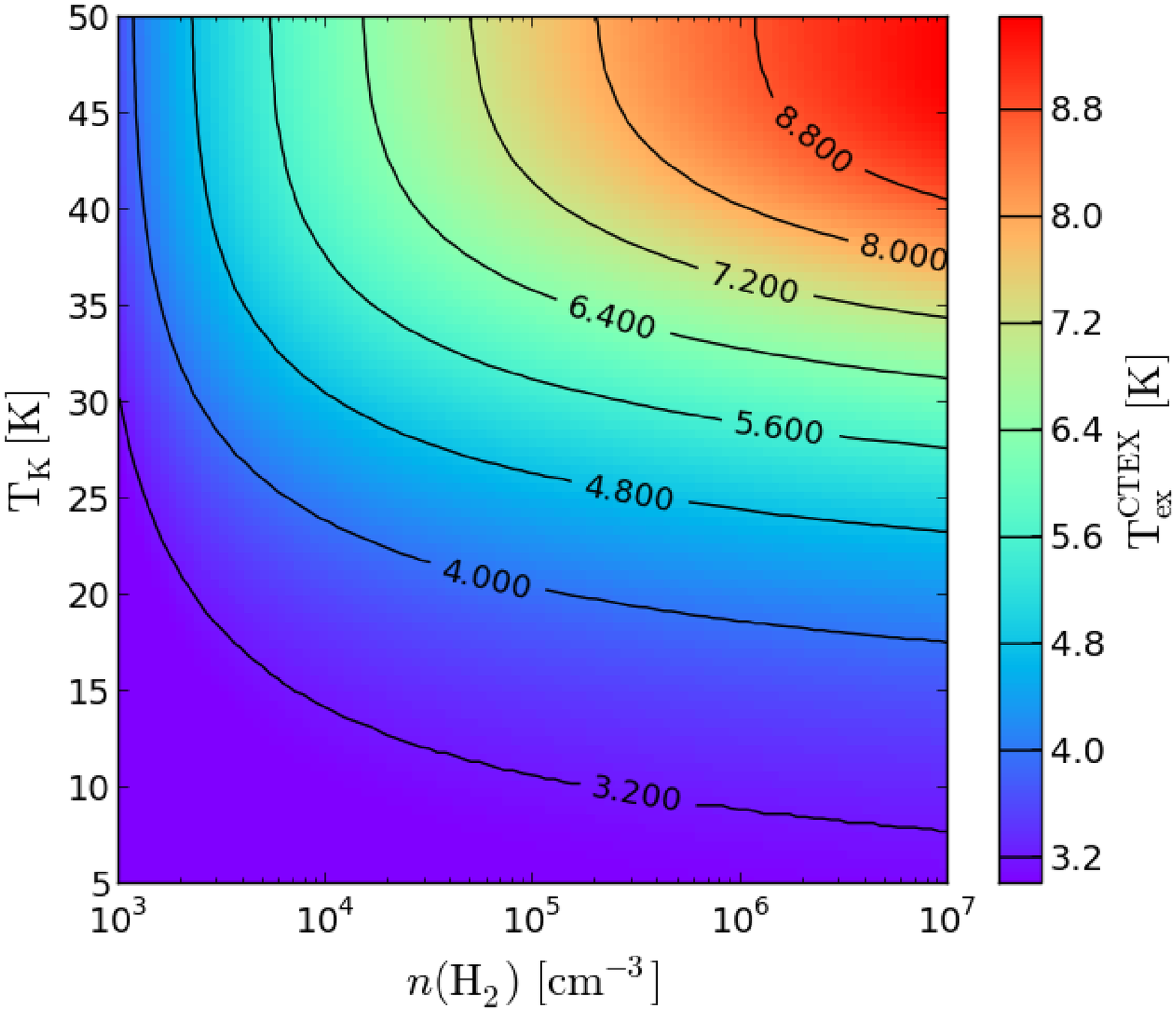}
\includegraphics{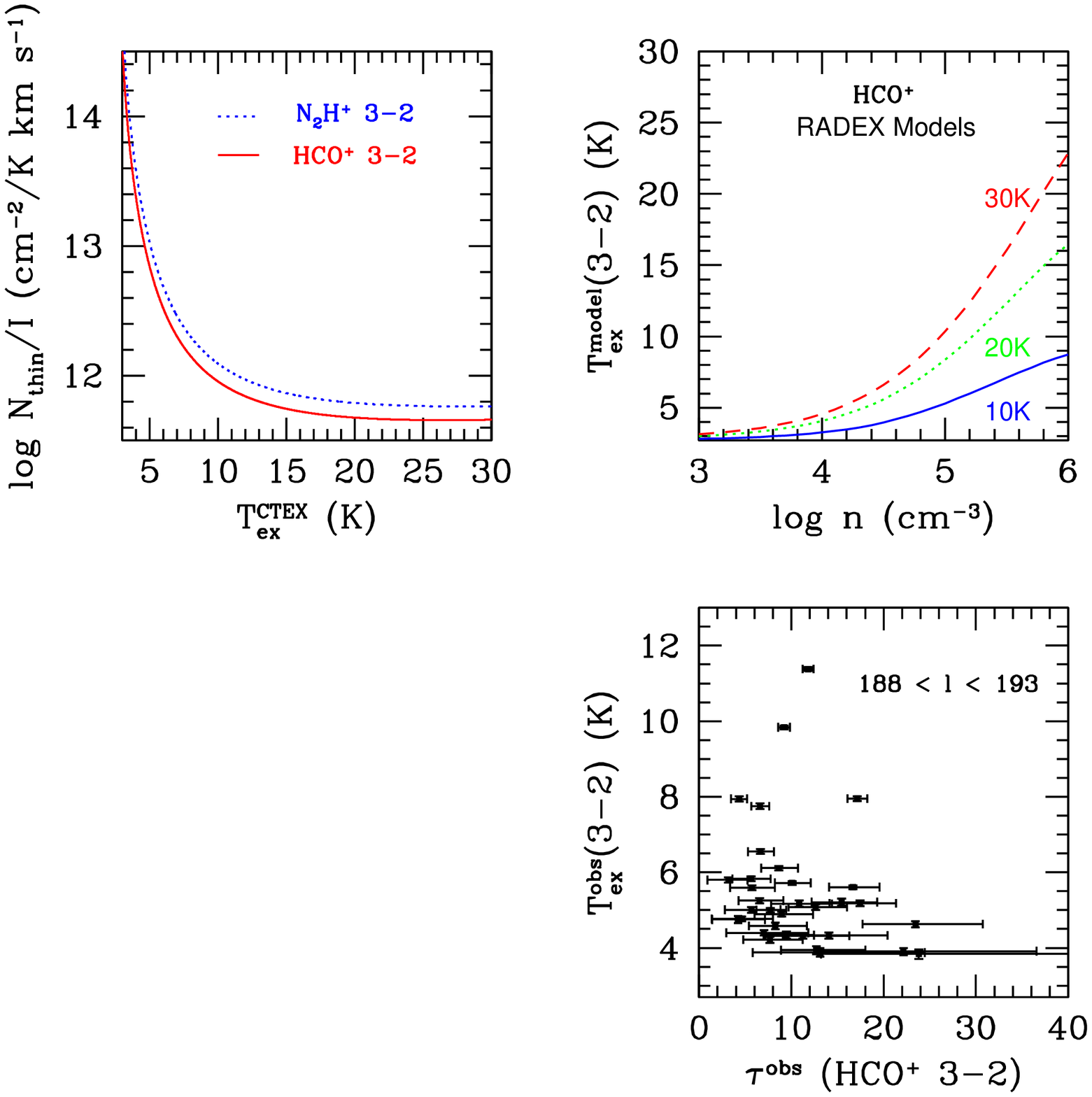}
\figcaption{TOP LEFT: Optically thin column density of \hcop\ $3-2$ (solid red line) 
and \nthp\ $3-2$ (dotted
blue line) versus $T_{ex}^{CTEX}$.  Below $T_{ex}^{CTEX} = 10$ K, 
the column density becomes a sensitive
function of excitation temperature.
TOP RIGHT: The model excitation temperature of \hcop\ $3-2$
versus volume density for constant density, constant kinetic temperature
radiative transfer models with $\log N/\Delta v = 13.5$ (RADEX; van der Tak et al. 2007).  
The blue (solid) curves is $T_k = 10$ K, the green (dotted) curve is $T_k = 20$ K, 
and the red (dashed) curve is $T_k = 30$ K.  
BOTTOM LEFT:  Contours of $T_{ex}^{CTEX}$ for \hcop\ $3-2$ observations
from constant density, constant kinetic temperature models.
BOTTOM RIGHT: Peak optical depth of the \hcop\ $3-2$ line calculated from \hcopi\ $3-2$ 
observations. \hcop\ $3-2$ emission is typically optically thick and sub-thermally
populated in the physical conditions found in BGPS clumps.}
\end{figure}


\begin{figure}
\figurenum{14}
\epsscale{0.9}
\plotone{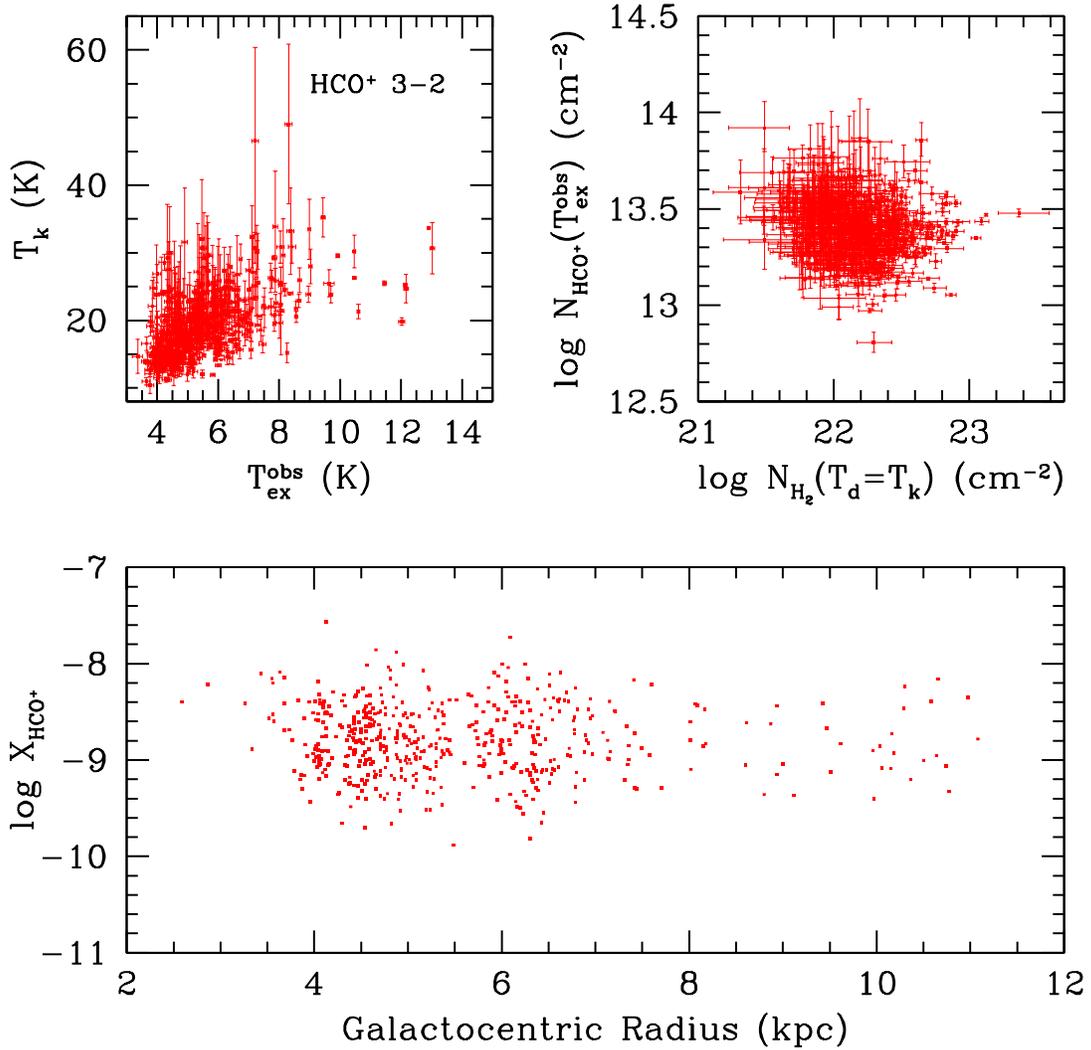}
\figcaption{TOP LEFT: The gas kinetic temperature versus the excitation temperature determined from the \hcop\ $3-2$ line.  TOP RIGHT: The column density of \hcop\ versus the H$_2$ column density assuming $T_d = T_k$.  Only the subset of sources with NH$_3$ observations
are plotted.
BOTTOM: The \hcop\ $3-2$ abundance does not vary systematically with Galactocentric radius.}
\end{figure}


\begin{deluxetable}{lccccc}
\footnotesize
\tablecolumns{6}
\tablecaption{Calibration Efficiencies \label{tab1}}
\tablewidth{0pt}
\tablehead{
\colhead{MJD Range} &
\colhead{$\eta_{LSB}^{Vpol}$\tablenotemark{a}} &
\colhead{$\eta_{LSB}^{Hpol}$} &
\colhead{$\eta_{USB}^{Vpol}$} &
\colhead{$\eta_{USB}^{Hpol}$} &
\colhead{Catalog\tablenotemark{b}}
}
\startdata
54863 - 54916	& 0.81 (0.04)	& 0.70 (0.03)	& 0.81 (0.03)	& 0.70 (0.04) & 1 \\
54917 - 54919	& 0.64 (0.01)	& 0.64 (0.01)	& 0.64 (0.02)	& 0.64 (0.02) & 1 \\
54920 - 54997	& 0.81 (0.04)	& 0.70 (0.03)	& 0.81 (0.03)	& 0.70 (0.04) & 1 \\
55620 - 55669	& 0.66 (0.03)	& 0.56 (0.03)	& 0.71 (0.03)	& 0.62 (0.03) & 2 \\
55678 - 56216	& 0.83 (0.03)	& 0.75 (0.03)	& 0.82 (0.03)	& 0.78 (0.03) & 2 \\
\enddata
\tablenotetext{a}{$\eta_{SB}^{pol} = \eta_{mb}\eta_{pol}$.}
\tablenotetext{b}{Catalog number 1 = Schlingman et al. (2011) and 2 = new observations presented 
in this paper.}
\end{deluxetable}

\begin{deluxetable}{llllcccccccccc}
\rotate
\tabletypesize{\scriptsize}
\tablecolumns{14}
\tablecaption{\hcop\ $3-2$ Observed Properties \label{tab2}}
\tablewidth{0pt}
\tablehead{
\colhead{Number} &
\colhead{Source} &
\colhead{$\alpha$(J2000.0)} &
\colhead{$\delta$} &
\colhead{Catalog\tablenotemark{a}} &
\colhead{Flag} &
\colhead{$v_{LSR}$} &
\colhead{$T_{mb}^{pk}$} &
\colhead{$\sigma_{T_{mb}}$} &
\colhead{$I(T_{mb})$} &
\colhead{$\sigma_I$} & 
\colhead{$\Delta v$} &
\colhead{$\sigma_{\Delta v}$} &
\colhead{FWZI} \\
\colhead{} &
\colhead{} &
\colhead{h:m:s} &
\colhead{d:m:s} &
\colhead{} &
\colhead{} &
\colhead{(km/s)} &
\colhead{(K)} &
\colhead{(K)} &
\colhead{(K km/s)} &
\colhead{(K km/s)} & 
\colhead{(km/s)} &
\colhead{(km/s)} &
\colhead{(km/s)} 
}
\startdata
1307 &        G007.501+00.001 & 18:02:30.0 & -22:28:07.0 &  2 & 0 & \nodata &  \nodata & 0.095 & \nodata & \nodata &  \nodata & \nodata &  \nodata  \\ 
1308 &        G007.507-00.255 & 18:03:28.7 & -22:35:22.3 &  2 & 0 & \nodata &  \nodata & 0.104 & \nodata & \nodata &  \nodata & \nodata &  \nodata  \\ 
1309 &        G007.509+00.403 & 18:01:00.4 & -22:15:46.3 &  2 & 1 & 7.5 &  0.450 & 0.086 & 1.245 & 0.288 &  2.7 & 0.5 &  \nodata  \\ 
1310 &        G007.564-00.042 & 18:02:47.8 & -22:26:05.6 &  2 & 0 & \nodata &  \nodata & 0.070 & \nodata & \nodata &  \nodata & \nodata &  \nodata  \\ 
1311 &        G007.600-00.142 & 18:03:15.0 & -22:27:10.1 &  2 & 1 & 153.3 &  0.477 & 0.061 & 1.384 & 0.235 &  3.4 & 0.4 &  3.4  \\ 
1312 &        G007.622+00.002 & 18:02:45.3 & -22:21:45.8 &  2 & 0 & \nodata &  \nodata & 0.081 & \nodata & \nodata &  \nodata & \nodata &  \nodata  \\ 
1313 &        G007.622-00.000 & 18:02:45.8 & -22:21:49.4 &  2 & 0 & \nodata &  \nodata & 0.080 & \nodata & \nodata &  \nodata & \nodata &  \nodata  \\ 
1314 &        G007.632-00.110 & 18:03:11.9 & -22:24:33.1 &  2 & 1 & 153.9 &  2.254 & 0.066 & 13.348 & 0.290 &  6.2 & 0.2 &  11.2  \\ 
1315 &        G007.636-00.150 & 18:03:21.4 & -22:25:31.4 &  2 & 1 & 154.8 &  0.519 & 0.060 & 1.412 & 0.200 &  2.7 & 0.3 &  3.4  \\ 
1316 &        G007.636-00.194 & 18:03:31.4 & -22:26:49.4 &  2 & 1 & 152.6 &  1.167 & 0.055 & 6.192 & 0.248 &  5.5 & 0.2 &  10.1  \\ 
\enddata
\tablenotetext{0}{Table 2 is published in its entirety in the electronic edition of the ApJS, a portion is shown here for guidance regarding its form and content.}
\tablenotetext{a}{Catalog number 1 = Schlingman et al. (2011) and 2 = new observations presented 
in this paper.}
\end{deluxetable}

\begin{deluxetable}{llcccccccccc}
\tabletypesize{\scriptsize}
\rotate
\tablecolumns{12}
\tablecaption{\nthp\ $3-2$ Observed Properties \label{tab3}}
\tablewidth{0pt}
\tablehead{
\colhead{Number} &
\colhead{Source} &
\colhead{Catalog\tablenotemark{a}} &
\colhead{Flag} &
\colhead{$v_{LSR}$} &
\colhead{$T_{mb}^{pk}$} &
\colhead{$\sigma_{T_{mb}}$} &
\colhead{$I(T_{mb})$} &
\colhead{$\sigma_I$} & 
\colhead{$\Delta v$} &
\colhead{$\sigma_{\Delta v}$} &
\colhead{FWZI} \\
\colhead{} &
\colhead{} &
\colhead{} &
\colhead{} &
\colhead{(km/s)} &
\colhead{(K)} &
\colhead{(K)} &
\colhead{(K km/s)} &
\colhead{(K km/s)} & 
\colhead{(km/s)} &
\colhead{(km/s)} &
\colhead{(km/s)} 
}
\startdata
1307 &        G007.501+00.001 &  2 & 0 & \nodata &  \nodata & 0.113 & \nodata & \nodata &  \nodata & \nodata &  \nodata  \\ 
1308 &        G007.507-00.255 &  2 & 0 & \nodata &  \nodata & 0.097 & \nodata & \nodata &  \nodata & \nodata &  \nodata  \\ 
1309 &        G007.509+00.403 &  2 & 0 & \nodata &  \nodata & 0.093 & \nodata & \nodata &  \nodata & \nodata &  \nodata  \\ 
1310 &        G007.564-00.042 &  2 & 0 & \nodata &  \nodata & 0.078 & \nodata & \nodata &  \nodata & \nodata &  \nodata  \\ 
1311 &        G007.600-00.142 &  2 & 1 & 153.0 &  0.404 & 0.082 & 1.979 & 0.308 &  \nodata & \nodata &  \nodata  \\ 
1312 &        G007.622+00.002 &  2 & 0 & \nodata &  \nodata & 0.067 & \nodata & \nodata &  \nodata & \nodata &  \nodata  \\ 
1313 &        G007.622-00.000 &  2 & 0 & \nodata &  \nodata & 0.078 & \nodata & \nodata &  \nodata & \nodata &  \nodata  \\ 
1314 &        G007.632-00.110 &  2 & 1 & 153.9 &  1.591 & 0.079 & 7.967 & 0.339 &  5.3 & 0.2 &  8.6  \\ 
1315 &        G007.636-00.150 &  2 & 1 & 155.3 &  0.661 & 0.068 & 1.633 & 0.224 &  2.5 & 0.3 &  4.3  \\ 
1316 &        G007.636-00.194 &  2 & 1 & 152.8 &  0.858 & 0.086 & 4.679 & 0.380 &  4.9 & 0.3 &  6.4  \\ 
\enddata
\tablenotetext{0}{Table 3 is published in its entirety in the electronic edition of the ApJS, a portion is shown here for guidance regarding its form and content.}
\tablenotetext{a}{Catalog number 1 = Schlingman et al. (2011) and 2 = new observations presented 
in this paper.}
\end{deluxetable}

\begin{deluxetable}{llcccc}
\footnotesize
\tablecolumns{6}
\tablecaption{\hcop\ $3-2$ Optical Depth from \hcopi\ 3-2 Observations in Gemini \label{tab4}}
\tablewidth{0pt}
\tablehead{
\colhead{Number} &
\colhead{Source} &
\colhead{$T_{mb}$(\hcopi )} &
\colhead{$\sigma_{T_{mb}}$} &
\colhead{$\tau$(\hcop )\tablenotemark{a}} &
\colhead{$\sigma_{\tau}$} \\
\colhead{} &
\colhead{} &
\colhead{(K)} &
\colhead{(K)} &
\colhead{} &
\colhead{} 
}
\startdata
7460 &        G188.792+01.027	&	0.363	&	0.043	&	6.61	&	0.98	\\
7461 &        G188.948+00.883	&	1.303	&	0.035	&	8.54	&	0.33	\\
7462 &        G188.975+00.911	&	0.208	&	0.047	&	10.84	&	3.32	\\
7463 &        G188.991+00.859	&	0.067	&	0.035	&	4.23	&	2.86	\\
7464 &        G189.015+00.823	&	0.225	&	0.040	&	12.64	&	3.16	\\
7465 &        G189.030+00.781	&	1.273	&	0.048	&	11.81	&	0.60	\\
7471 &        G189.682+00.185	&	0.316	&	0.040	&	17.44	&	3.60	\\
7472 &        G189.713+00.335 	&	0.136	&	0.040	&	23.81	&	14.90	\\
7473 &        G189.744+00.335	&	0.110	&	0.031	&	8.26	&	3.16	\\
7474 &        G189.776+00.343	&	0.897	&	0.033	&	17.13	&	1.08	\\
7475 &        G189.782+00.265 	&	0.280	&	0.044	&	23.45	&	6.53	\\
7476 &        G189.782+00.323	&	\nodata	&	0.040	&	$<$ 6.74	& \nodata		\\
7477 &        G189.783+00.433	&	\nodata	&	0.036	&	$<$ 12.17	& \nodata		\\
7478 &        G189.783+00.465	&	\nodata	&	0.047	&	$<$ 34.5	& \nodata		\\
7479 &        G189.788+00.281	&	0.102	&	0.019	&	9.45	&	2.75	\\
7480 &        G189.789+00.291 	&	0.082	&	0.042	&	7.03	&	4.43	\\
7481 &        G189.804+00.355	&	0.259	&	0.038	&	10.09	&	1.93	\\
7482 &        G189.810+00.369 	&	0.137	&	0.041	&	6.55	&	2.43	\\
7483 &        G189.831+00.343	&	0.138	&	0.021	&	7.73	&	1.61	\\
7484 &        G189.834+00.317 	&	0.147	&	0.041	&	8.96	&	3.17	\\
7485 &        G189.836+00.303	&	\nodata	&	0.033	&	$<$ 14.69	& \nodata		\\
7486 &        G189.864+00.499 	&	0.095	&	0.049	&	3.18	&	2.18	\\
7487 &        G189.879+00.319	&	\nodata	&	0.041	&	$<$ 24.86	& \nodata		\\
7488 &        G189.885+00.319	&	0.090	&	0.017	&	12.71	&	4.55	\\
7489 &        G189.888+00.303	&	0.118	&	0.035	&	11.26	&	4.56	\\
7490 &        G189.921+00.331 	&	\nodata	&	0.041	&	$<$ 8.11	& \nodata		\\
7491 &        G189.950+00.231	&	0.104	&	0.045	&	5.67	&	3.00	\\
7492 &        G189.951+00.331	&	0.382	&	0.042	&	16.68	&	2.71	\\
7493 &        G189.990+00.353	&	\nodata	&	0.048	&	$<$ 50.30	& \nodata		\\
7494 &        G190.006+00.361	&	0.140	&	0.041	&	22.16	&	11.79	\\
7495 &        G190.044+00.543	&	\nodata	&	0.045	&	$<$ 21.24	& \nodata		\\
7497 &        G190.063+00.679 	&	\nodata	&	0.043	&	$<$ 37.16	& \nodata		\\
7498 &        G190.171+00.733	&	0.289	&	0.042	&	15.45	&	3.56	\\
7499 &        G190.192+00.719 	&	0.087	&	0.040	&	13.12	&	9.33	\\
7500 &        G190.240+00.911	&	0.076	&	0.023	&	7.67	&	3.22	\\
7501 &        G192.581-00.043	&	0.790	&	0.042	&	9.22	&	0.64	\\
7502 &        G192.596-00.051	&	0.260	&	0.042	&	4.34	&	0.86	\\
7503 &        G192.602-00.143	&	\nodata	&	0.045	&	$<$ 18.08	& \nodata		\\
7505 &        G192.629-00.157	&	0.144	&	0.040	&	14.07	&	5.69	\\
7506 &        G192.644+00.003 	&	0.074	&	0.043	&	4.64	&	3.32	\\
7507 &        G192.662-00.083	&	\nodata	&	0.038	&	$<$ 20.80	& \nodata		\\
7508 &        G192.719+00.043	&	\nodata	&	0.048	&	$<$ 9.60	& \nodata		\\
7509 &        G192.764+00.101	&	0.160	&	0.050	&	5.63	&	2.08	\\
7510 &        G192.816+00.127 	&	\nodata	&	0.022	&	$<$ 5.73	& \nodata		\\
7511 &        G192.968+00.093	&	\nodata	&	0.051	&	$<$ 10.03	& \nodata		\\
7512 &        G192.981+00.149	&	0.250	&	0.043	&	6.66	&	1.40	\\
7513 &        G192.985+00.177	&	0.145	&	0.052	&	5.72	&	2.43	\\
7514 &        G193.006+00.115	&	0.268	&	0.050	&	8.66	&	2.00	\\
\enddata
\tablenotetext{a}{Upper limits correspond to $3\sigma_{T_{mb}}$ upper limits for \hcopi\ $3-2$.}
\end{deluxetable}

\begin{deluxetable}{llcccc}
\footnotesize
\tablecolumns{6}
\tablecaption{\hcop\ $3-2$ Line Asymmetries \label{tab5}}
\tablewidth{0pt}
\tablehead{
\colhead{Number} &
\colhead{Source} &
\colhead{Catalog\tablenotemark{a}} &
\colhead{\hcop\ Flag} &
\colhead{\nthp\ Flag} &
\colhead{Asymmetry\tablenotemark{b}} 
}
\startdata
1363 &        G008.458-00.224 &  2 & 3 & 1 & b  \\ 
1377 &        G008.670-00.356 &  2 & 3 & 1 & b  \\ 
1398 &        G008.872-00.318 &  2 & 3 & 1 & r  \\ 
1412 &        G009.212-00.202 &  2 & 3 & 1 & b  \\ 
1421 &        G009.620+00.194 &  2 & 3 & 1 & r  \\ 
1466 &        G010.214-00.324 &  1 & 3 & 3 & b  \\ 
1491 &        G010.416-00.030 &  2 & 3 & 0 & b  \\ 
1518 &        G010.681-00.028 &  1 & 3 & 1 & b  \\ 
1521 &        G010.693-00.404 &  1 & 3 & 0 & b  \\ 
1584 &        G011.083-00.536 &  1 & 3 & 1 & b  \\ 
1659 &        G011.947-00.036 &  1 & 3 & 1 & r  \\ 
1780 &        G012.809-00.200 &  1 & 3 & 1 & r  \\ 
1796 &        G012.861-00.272 &  1 & 3 & 1 & r  \\ 
1803 &        G012.889+00.490 &  1 & 3 & 1 & b  \\ 
1833 &        G012.999-00.358 &  1 & 3 & 1 & r  \\ 
1869 &        G013.211-00.142 &  1 & 3 & 1 & r  \\ 
1943 &        G013.816+00.003 &  2 & 3 & 0 & b  \\ 
1956 &        G013.882-00.143 &  2 & 3 & 1 & b  \\ 
2019 &        G014.244-00.071 &  1 & 3 & 1 & e  \\ 
\enddata
\tablenotetext{0}{Table 4 is published in its entirety in the electronic edition of the ApJS, a portion is shown here for guidance regarding its form and content.}
\tablenotetext{a}{Catalog number 1 = Schlingman et al. (2011) and 2 = new observations presented 
in this paper.}
\tablenotetext{b}{b = blue-asymmetry, r = red asymmetric, e = equal asymmetry.}
\end{deluxetable}

\end{document}